\documentclass[%
onecolumn,
 amsmath,amssymb,
 aps,prfluids
]{revtex4-2}

\usepackage{graphicx}
\usepackage{dcolumn}
\usepackage{bm}
\usepackage{hyperref}

\begin{document}

\preprint{APS/123-QED}

\title{Robust wall modes and their interplay with bulk turbulence in confined rotating Rayleigh-B\'{e}nard convection}

\author{Xander M. de Wit}
\author{Wouter J. M. Boot}
\author{Matteo Madonia}
\author{Andr\'{e}s J. Aguirre Guzm\'{a}n}
\author{Rudie P. J. Kunnen}%
 \email{r.p.j.kunnen@tue.nl}
\affiliation{%
Fluids and Flows group, Department of Applied Physics and J. M. Burgers Centre for Fluid Dynamics, Eindhoven University of Technology, P.O. Box 513, 5600 MB Eindhoven, Netherlands
}%


\date{25 July, 2023}

\begin{abstract}
In confined rotating convection, a strong zonal flow can develop close to the side wall with a modal structure that precesses anti-cyclonically (counter to the applied rotation) along the side wall. It is surmised that this is a robust non-linear evolution of the wall modes observed before the onset of bulk convection. Here, we perform direct numerical simulations of cylindrically confined rotating convection at high rotation rates and strong turbulent forcing. Through comparison with earlier work, we find a fit-parameter-free relation that links the angular drift frequency of the robust wall mode observed far into the turbulent regime with the critical wall mode frequency at onset, firmly substantiating the connection between the observed boundary zonal flow and the wall modes. Deviations from this relation at stronger turbulent forcing suggest early signs of the bulk turbulence starting to hamper the development of the wall mode. Furthermore, by studying the interactive flow between the robust wall mode and the bulk turbulence, we identify radial jets penetrating from the wall mode into the bulk. These jets induce a large scale multipolar vortex structure in the bulk turbulence, dependent on the wavenumber of the wall mode. In a narrow cylinder the entire bulk flow is dominated by a quadrupolar vortex driven by the radial jets, while in a wider cylinder the jets are found to have a finite penetration length and the vortices do not cover the entire bulk. We also identify the role of Reynolds stresses in the generation of zonal flows in the region near the sidewall.
\end{abstract}

\maketitle


\section{Introduction}
Rotating Rayleigh-B\'{e}nard convection is the principal source of motion underlying the majority of geophysical and astrophysical flows \cite{Aurnou2015,Miesch2000,Heimpel2005}. There, a turbulent flow is driven by buoyancy and is simultaneously affected by background rotation, typically the celestial rotation. It is fundamental to many oceanic and atmospheric flows as well as for example solar convection, flows in the Earth's liquid metal core and on gas giant planets such as Jupiter and Saturn. This enticingly rudimentary flow set-up encompasses a rich phenomenology of different flow regimes \cite{Ecke2023,Kunnen2021}. Investigations are continuing to advance to more extreme parameters of strong turbulent forcing and rotation in order to come closer to the flows encountered in nature, pushing the boundaries for both laboratory investigations \cite{Bouillaut2021,Madonia2021,Wedi2021} as well as numerical simulations \cite{Stellmach2014,Maffei2021,AguirreGuzman2022}.

While numerical simulations allow to study laterally unbounded convection by employing periodic boundary conditions on the sides of the domain, laboratory investigations of rotating Rayleigh-B\'{e}nard convection inevitably need to resort to confined domains, typically using a cylindrical tank. The consequences of this lateral confinement thus need to be well understood in order to draw comparisons with the large scale convective flows observed in nature. Recently, it has been found that in confined rotating convection, a strong zonal flow can emerge in a region close to the sidewall of the flow domain \cite{DeWit2020,Zhang2020,Favier2020,Shishkina2020,Zhang2021,Lu2021,Wedi2022,Terrien2023}, which was termed the boundary zonal flow (BZF), sidewall circulation, or wall mode. This flow structure is comprised of alternating sections of hot rising fluid and cold sinking fluid, carrying a large convective heat flux, and it precesses anti-cyclonically along the sidewall. It is conjectured that this flow structure is related to the wall modes observed before the onset of bulk convection \cite{Favier2020,Ecke2022}. Surprisingly, this wall mode state seems to remain robust in its non-linear evolution well into the turbulent regime of the flow. Moreover, jet-like bursts emanating from the wall mode have been observed \cite{Favier2020,Madonia2021,Madonia2023}, originating from the positions where hot rising fluid and cold sinking fluid meet, that have a profound effect on the overall circulation in the bulk.

In this work, we consider direct numerical simulations (DNSs) of cylindrically confined rotating Rayleigh-B\'{e}nard convection to further investigate the properties of these robust wall modes and their interaction with the bulk turbulence. The numerical method as well as the employed parameters and resolutions are provided in Section~\ref{sec:numerical}. By relating our simulations to results in other recent work, we find a scaling relation for the precession frequency of the robust wall mode that firmly establishes the surmised relation between the BZF observed in the turbulent regime and the wall mode state found before the onset of bulk convection as laid out in Section~\ref{sec:wallmode}. We then show in Section~\ref{sec:interaction} how jets emerging from the wall mode can induce a large multipolar vortex in the bulk. This large vortex shows morphological similarities with the large scale vortices (LSVs) observed in unconfined convection, originating from an upscale flux of kinetic energy in a similar parameter range \cite{Favier2014,Guervilly2014,Guzman2020,Julien2012,Rubio2014,Stellmach2014,DeWit2022}. However, by considering a wider cylindrical geometry, we show that this multipolar vortex is connected to the wavenumber of the wall mode, suggesting that this vortex emerges predominantly as a consequence of the wall mode to bulk interaction rather than originating from an upscale flux of turbulent kinetic energy. Finally, in Section~\ref{sec:RANS}, we consider the Reynolds-averaged properties of the wall modes by studying the full balance between different terms in the Reynolds-average Navier Stokes (RANS) equations, revealing the source terms that drive the observed mean zonal flow in the wall mode. The conclusions and outlook are provided in Section~\ref{sec:conclusion}.

\section{Numerical method}\label{sec:numerical}
Rotating Rayleigh-B\'{e}nard convection is governed by three independent dimensionless numbers, quantifying the strength of the buoyant forcing, the fluid properties and (inverse) strength of rotation as represented by, respectively, the Rayleigh number $\textrm{Ra}$, the Prandtl number $\textrm{Pr}$ and the Ekman number $\textrm{Ek}$ as
\begin{equation}
    \textrm{Ra}=\frac{g\alpha\Delta T H^3}{\nu\kappa}, \qquad \textrm{Pr}=\frac{\nu}{\kappa},\qquad \textrm{Ek}=\frac{\nu}{2\Omega H^2},
\end{equation}
where $g$ is the gravitational acceleration, the fluid properties $\alpha$, $\nu$ and $\kappa$ respectively represent the thermal expansion coefficient, kinematic viscosity and thermal diffusivity of the working fluid, $\Delta T$ denotes the temperature difference between the hot bottom and cold top of the flow domain, while $H$ is the domain height and $\Omega$ is the angular velocity of the background rotation, assumed antiparallel to gravity. In literature, also the convective Rossby number $\textrm{Ro}=\textrm{Ek}(\textrm{Ra}/\textrm{Pr})^{1/2}$ is often used. Additionally, in confined rotating convection, the diameter-to-height aspect ratio $\Gamma=2R/H$ of the domain plays a role.

The flow is governed by the Navier-Stokes and heat equations for an incompressible Boussinesq fluid. In dimensionless form, this yields \cite{Chandrasekhar1961}
\begin{subequations}\begin{align}
    \frac{\partial\tilde{\boldsymbol{u}}}{\partial\tilde{t}} + \left(\tilde{\boldsymbol{u}}\cdot\tilde{\boldsymbol{\nabla}}\right)\tilde{\boldsymbol{u}} + \frac{1}{\textrm{Ro}}\boldsymbol{e}_z\times\tilde{\boldsymbol{u}} &= -\tilde{\boldsymbol{\nabla}}\tilde{p} + \left(\frac{\textrm{Pr}}{\textrm{Ra}}\right)^{1/2}\tilde{\nabla}^2\tilde{\boldsymbol{u}} + \tilde{T}\boldsymbol{e}_z, \label{eq:NS_momentum}\\
    \frac{\partial \tilde{T}}{\partial\tilde{t}} + \left(\tilde{\boldsymbol{u}}\cdot\tilde{\boldsymbol{\nabla}}\right)\tilde{T} &= \frac{1}{\left(\textrm{Ra}\textrm{Pr}\right)^{1/2}}\tilde{\nabla}^2\tilde{T}, \label{eq:NS_temp}\\
    \tilde{\boldsymbol{\nabla}}\cdot\tilde{\boldsymbol{u}} &= 0, \label{eq:NS_cont}
\end{align}\end{subequations}
describing the evolution of the flow field $\boldsymbol{u}$ and temperature field $T$ in time $t$, where $p$ is the pressure and $\boldsymbol{e}_z$ is the vertical unit vector. Here, tildes denote non-dimensionalization using the free-fall velocity $U=\sqrt{g\alpha\Delta T H}$ as the velocity scale, $H$ as the length scale and $\Delta T$ as the temperature scale.

\begin{table}[b]
\caption{\label{tab:input}Input parameters and resolutions that are used for the simulations in this work.}
\begin{ruledtabular}
\begin{tabular}{llllllll}
&\textrm{Ra}&\textrm{Pr}&\textrm{Ek}&\textrm{Ro}&$\Gamma$&$N_r\times N_\theta\times N_z$&\\
\colrule
Small aspect ratio \rule{0pt}{10pt} & $5.0\times10^{10}$ & 5.2 & $10^{-7}$ & $9.8\times10^{-3}$ & 0.20 & $351\times769\times1025$ &\footnote{covered in de Wit \textit{et al.} \cite{DeWit2020}}\\
& $7.0\times10^{10}$ & & & $1.2\times10^{-2}$ & & &$^{\textrm{a}}$\\
& $9.9\times10^{10}$ & & & $1.4\times10^{-2}$ & & &\\
& $1.4\times10^{11}$ & & & $1.6\times10^{-2}$ & & &$^{\textrm{a}}$\\
& $2.1\times10^{11}$ & & & $2.0\times10^{-2}$ & & & \\
& $3.2\times10^{11}$ & & & $2.5\times10^{-2}$ & & &$^{\textrm{a}}$\\
& $4.3\times10^{11}$ & & & $2.9\times10^{-2}$ & & &$^{\textrm{a}}$\\
& $6.0\times10^{11}$ & & & $3.4\times10^{-2}$ & & $469\times1025\times1365$ &\\
& $9.5\times10^{11}$ & & & $4.3\times10^{-2}$ & & &\\
& $1.5\times10^{12}$ & & & $5.4\times10^{-2}$ & & &\\
Larger aspect ratio & $2.0\times10^{11}$ & & & $2.0\times10^{-2}$ & 0.72 & $513\times2049\times1025$&
\end{tabular}
\end{ruledtabular}
\end{table}

The governing equations are solved numerically in a cylindrical coordinate system $(r,\theta,z)$ using a second order finite-difference code developed by Verzicco \& Orlandi \cite{Verzicco1996}. We employ no-slip constant temperature boundary conditions on both top and bottom, while the sidewalls of the cylinder are no-slip and fully insulating. Our simulations are carried out at constant $\textrm{Ek}=10^{-7}$ and $\textrm{Pr}=5.2$, representing water, extending our earlier set of simulations in Ref.~\cite{DeWit2020} to larger $\textrm{Ra}$. We use a slender cylinder with aspect ratio $\Gamma=0.20$ for the series at varying $\textrm{Ra}$, while one simulation is conducted in a wider cylinder at $\Gamma=0.72$, such that it fits two full wavelengths of the wall mode, see Section~\ref{sec:wallmode}. The full set of input parameters as well as the employed resolutions ($N_r\times N_\theta\times N_z$) are provided in Table~\ref{tab:input}.

The adequacy of the employed bulk resolution is validated \textit{a posteriori} through comparison with the local smallest dynamical length scale, i.e. the Batchelor scale $\eta_T$ (since we treat $\textrm{Pr}=5.2>1$). We ensure that in all directions, the grid spacing does not exceed $4\eta_T$, in agreement with the condition in Ref.~\cite{Verzicco2003}, although for the highest $\textrm{Ra}$ cases for both resolutions with $\Gamma=0.20$, this conditions needed to be relaxed to $5\eta_T$. To properly resolve the boundary layers near the top and bottom of the domain, we ensure that there are at least 10 grid cells in the Ekman boundary layers, complying with Ref.~\cite{Verzicco2003}.

\section{Wall mode: from onset to geostrophic turbulence}\label{sec:wallmode}
In confined rotating Rayleigh-B\'{e}nard convection, before the onset of bulk convection, the wall mode state emerges once a critical Rayleigh number $\textrm{Ra}_w$ is surpassed. Directly at onset of the wall mode, it starts to precess anti-cyclonically, i.e. counter to the direction of background rotation, at a nonzero frequency $\omega_{d_c}$. This critical point depends on $\textrm{Ek}$ (and $\textrm{Pr}$ in the case of $\omega_{d_c}$) and has been obtained from stability analysis as \cite{Herrmann1993,Zhang2009,Zhang2017}
\begin{equation}\label{eq:wall_critical}
    \textrm{Ra}_w \approx \pi^2 (6\sqrt{3})^{1/2} \textrm{Ek}^{-1} + 46.5 \textrm{ } \textrm{Ek}^{-2/3}, \qquad \omega_{d_c}/\Omega \approx (4 \pi^2 [3(2+\sqrt{3})]^{1/2} \textrm{Ek} - 1465 \textrm{ } \textrm{Ek}^{4/3})\textrm{Pr}^{-1}.
\end{equation}
The wall mode manifests as a wave adjacent to the sidewall of alternating hot rising fluid and cold sinking fluid, where, in the limit $\textrm{Ek}\to 0$, the azimuthal wavenumber $m$ is found to be \cite{Zhang2009,Zhang2017}
\begin{equation}\label{eq:wavenumber}
    m=\left[\frac{\pi}{2}(2+\sqrt{3})^{1/2}-17.49 \textrm{ } \textrm{Ek}^{1/3}\right]\Gamma.
\end{equation}
The recently observed BZF \cite{DeWit2020,Zhang2020}, encountered far beyond the onset of bulk convection, shows great morphological similarity with the wall mode state and has been surmised to be a long-lived non-linear evolution of the wall mode itself \cite{Favier2020,Ecke2022}.

\begin{figure}[h!]
\includegraphics[width=0.8\textwidth]{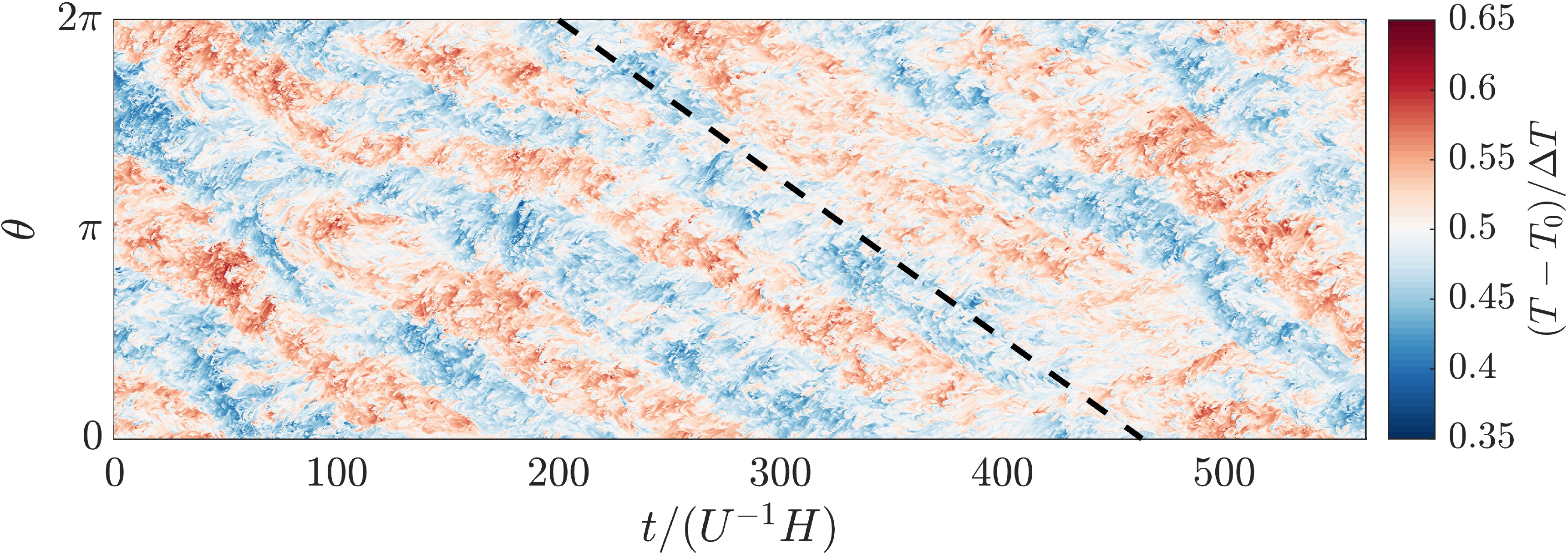}
\caption{\label{fig:moar_angle_time} Angle-time representation of the temperature of the wall mode, obtained at mid-height and $r=0.985 R$ for the case with $\textrm{Ra}=2.0\times10^{11}$ and $\Gamma=0.72$. The slope of the dashed line depicts the average angular drift velocity of the wall mode $\omega_d$.}
\end{figure}

The wall mode can be well appreciated from angle-time plots cross-sectioning the near wall region. An example for our wider cylinder case is provided in Fig.~\ref{fig:moar_angle_time}. From the figure, some slight defects are visible in the wall mode pattern, e.g. between $t \in [100,250] U^{-1}H$, deviating from a purely precessing $m=2$ mode. As we will argue later, we attribute this to the bulk turbulence starting the hamper the evolution of the wall mode.

From such angle-time data, the angular drift velocity of the wall mode can be determined by fitting the time series of azimuthal profiles of temperature (or vertical velocity) with sinusoids and tracking its phase \cite{DeWit2020,Favier2020,Zhang2021}. It has been shown that the onset of wall mode convection is a Hopf bifurcation where the wall mode sets in supercritically as a drifting wave \cite{Ecke1992,Goldstein1993}. These authors also show that the drift frequency $\omega_d$ depends linearly on the reduced bifurcation parameter $\varepsilon=(\textrm{Ra}-\textrm{Ra}_w)/\textrm{Ra}_w$ with a finite intercept $\omega_{d_c}$. In other words:
\begin{equation}\label{eq:ang_vel_prop}
    \omega_d - \omega_{d_c} \propto \textrm{Ra}-\textrm{Ra}_w.
\end{equation}
Remarkably, this relation has been found to hold not just close to onset, but even far beyond \cite{Favier2020,Ecke2022}. However, the prefactor, and hence also the functional dependence on the other parameters of the system ($\textrm{Ek}$, $\textrm{Pr}$ and $\Gamma$), has thus far only been approximated empirically \cite{Zhang2021}.

Here, we make an hypothesis for the prefactor by postulating that the same scaling of the critical point itself continues super-critically, i.e. that $\omega_{d_c}/\textrm{Ra}_w = (\omega_d - \omega_{d_c})/(\textrm{Ra}-\textrm{Ra}_w)$, yielding the relation
\begin{equation}\label{eq:ang_vel_scaling}
    \omega_d = \frac{\omega_{d_c}}{\textrm{Ra}_w} \textrm{Ra},
\end{equation}
where $\textrm{Ra}_w$ and $\omega_{d_c}$ are given by Eq.~\eqref{eq:wall_critical}. Note that this is consistent with Eq.~\eqref{eq:ang_vel_prop}, indeed, it follows equivalently from the hypothesis that both the scaling $\omega_d - \omega_{d_c} \propto \textrm{Ra}-\textrm{Ra}_w$ as well as $\omega_d \propto \textrm{Ra}$ hold simultaneously for all $\textrm{Ra}$. We emphasize that this relation is entirely fit-parameter-free: it has no remaining free coefficients in its dependence on system parameters and/or prefactors, but it is fully determined by the theoretical results from the asymptotic expansion, Eq.~\eqref{eq:wall_critical}.

\begin{figure}[h!]
\includegraphics[width=0.65\textwidth]{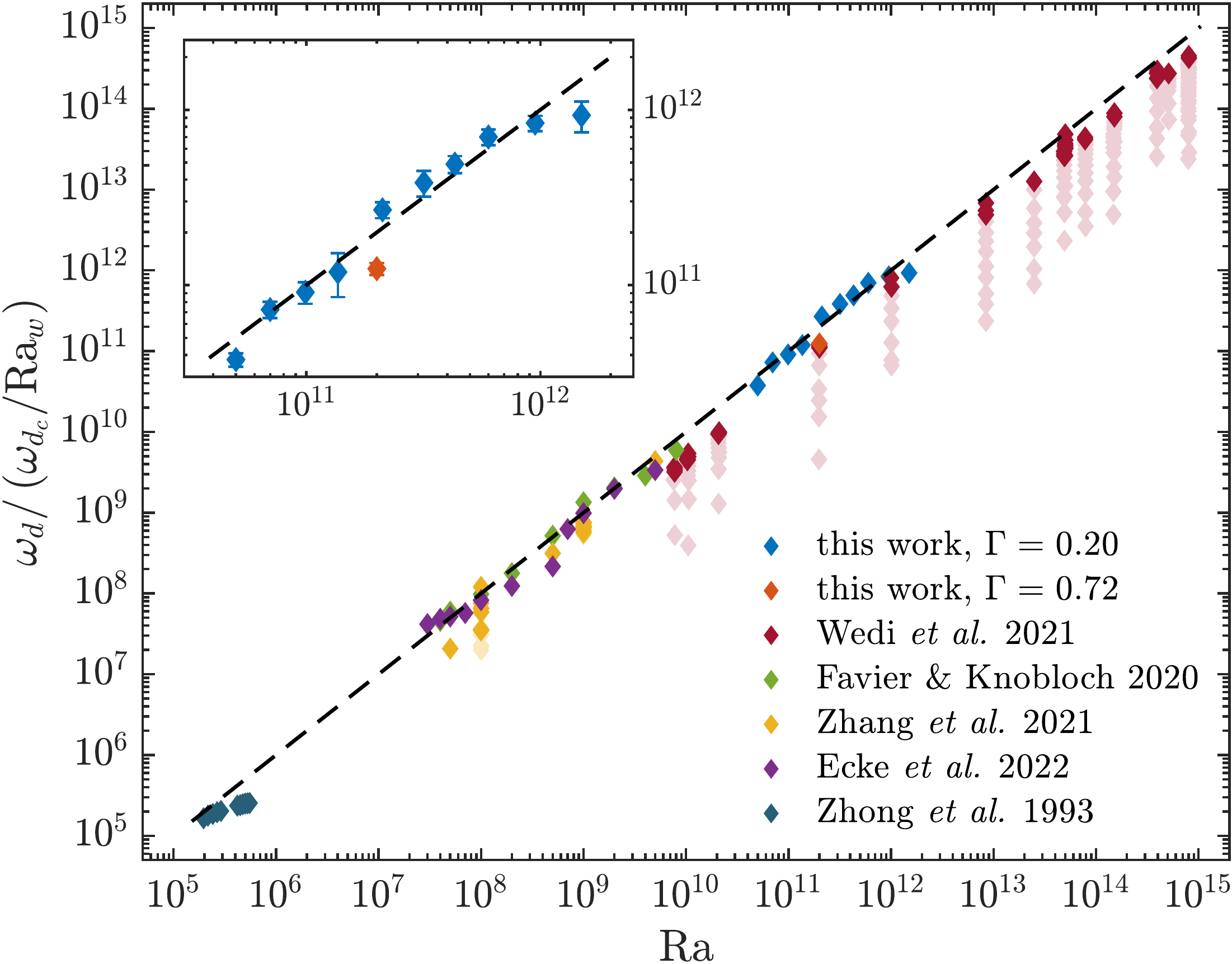}
\caption{\label{fig:ang_vel_all}Scaling of the angular drift velocity $\omega_d$ of the wall mode as a function of $\textrm{Ra}$ for our data and earlier works \cite{Zhang2021,Favier2020,Zhong1993,Ecke2022,Wedi2021}, following the fit-parameter-free relation Eq.~\eqref{eq:ang_vel_scaling}. For the data from Wedi \textit{et al.} \cite{Wedi2021} and Zhang \textit{et al.} \cite{Zhang2021}, data points with $\textrm{Ek}\textrm{Ra}^{1/2}>0.15$ are plotted in a lighter shade. The inset shows a close-up of our data including error bars.}
\end{figure}

Comparing this relation Eq.~\eqref{eq:ang_vel_scaling} to our data and that of earlier works, Fig.~\ref{fig:ang_vel_all}, we find very satisfactory agreement over ten decades in $\textrm{Ra}$. Nonetheless, deviations from this relation are also evident. We argue that Eq.~\eqref{eq:ang_vel_scaling} represents the `rotation-dominated scaling' of the drift velocity, where the wall mode remains fully developed. It is to be expected, however, that as the intensity of the bulk turbulence keeps increasing, relative to the strength of the wall mode, it will start to break down the wall mode at a certain stage as the flow loses rotational constraint. We interpret the deviations from Eq.~\eqref{eq:ang_vel_scaling} as first signatures of this break down of the wall mode. As shown in the figure in a lighter shade, based on the data from Wedi \textit{et al.}~\cite{Wedi2021}, this break down starts when a certain value of $\textrm{Ek}\textrm{Ra}^{1/2}$ is exceeded (that is, similar to a convective Rossby number $\textrm{Ro}$) although the dependence on $\textrm{Pr}$ and $\Gamma$ can not be determined from this data as these are kept constant. As shown in the inset of Fig.~\ref{fig:ang_vel_all}, also our simulations show a first sign of this hypothesized break down, with the drift velocity plateauing at the highest considered $\textrm{Ra}$ cases. Moreover, our data suggests that the point at which this break down of the wall mode commences also depends on the aspect ratio $\Gamma$, since small but significant deviations from Eq.~\eqref{eq:ang_vel_scaling} can be observed in the simulation with larger $\Gamma$. Qualitatively, one can argue that here, the spatially larger bulk turbulence hampers the wall mode more strongly.

In the limit $\textrm{Ek}\to0$, to leading order in $\textrm{Ek}$, the relation Eq.~\eqref{eq:ang_vel_scaling} becomes
\begin{equation}\label{eq:ang_vel_scaling_limit}
    \omega_d / \Omega = \frac{2(1+\sqrt{3})}{3^{1/4}} \textrm{Ek}^{2}\textrm{Pr}^{-1}\textrm{Ra},
\end{equation}
which is close to the exponents obtained empirically by Zhang \textit{et al.}~\cite{Zhang2021}, who found $\omega_d/\Omega \propto \textrm{Ek}^{5/3} \textrm{Pr}^{-4/3}\textrm{Ra}$. We compare the two scaling relations in some more detail in Appendix~\ref{app:scaling_alt}.

\section{Interaction between wall mode and bulk}\label{sec:interaction}
\subsection{Small aspect ratio: the quadrupolar vortex}
In this section, we study the flow patterns that emerge in the interaction between the wall mode and the bulk turbulence. To that extent, we resort to orientation compensated averages of the flow. These are constructed by tracking the phase angle of the wall mode. Then each snapshot of the flow is rotated back by this angle and the average is taken over time, see also Ref.~\cite{DeWit2020}.

A typical example of the instantaneous flow and the orientation compensated mean flow for one of the cases with $\Gamma=0.20$ is shown in Fig.~\ref{fig:small_together}, while individual velocity components and temperature of the mean flow are provided in Appendix~\ref{app:small_phase_ave}. This reveals the formation of a domain-spanning quadrupolar vortex in the bulk. This quadrupolar vortex is aligned with the orientation of the wall mode, such that jets emerging from the wall mode where the hot rising section meets the cold sinking section are feeding this quadrupolar vortex.

There is an evident asymmetry between the strength of the cyclonic and anti-cyclonic poles of the vortex. The anti-cyclonic poles are relatively stronger, which we attribute to the Coriolis effect deflecting the jets emerging from the wall region disproportionally to the right towards the anti-cyclone.

\begin{figure}[h!]
\includegraphics[width=\textwidth]{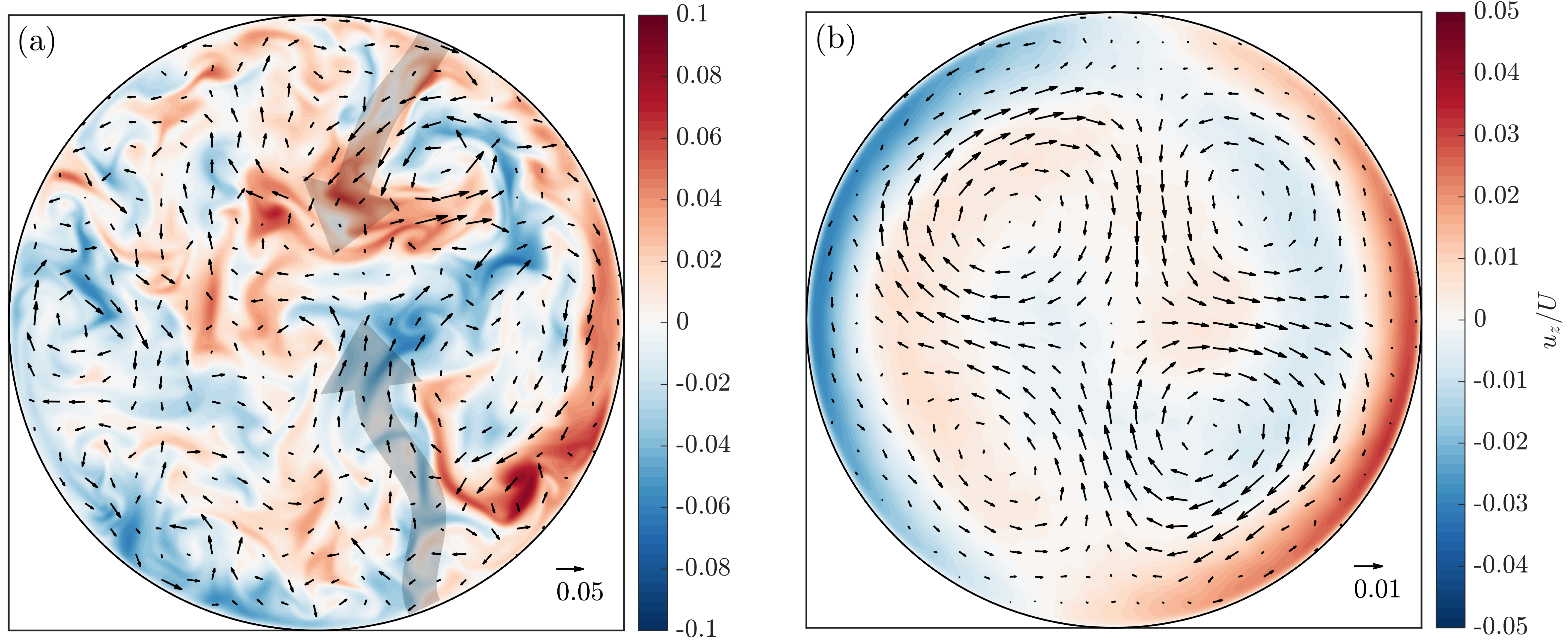}
\caption{\label{fig:small_together}A snapshot of the instantaneous flow (a) and the orientation compensated mean flow (b) at mid-height for the case $\textrm{Ra}=9.5\times10^{11}$ and $\Gamma=0.20$, showing the emergence of jets and the quadrupolar vortex in the bulk. Arrows indicate horizontal in-plane velocities and color depicts vertical velocity. Shaded arrows are used in (a) to highlight the jets originating from the wall mode.}
\end{figure}

The vortex structure observed here shares many characteristics with the large scale vortices (LSVs) observed in laterally unbounded rotating convection \cite{Favier2014,Guervilly2014,Guzman2020,Julien2012,Rubio2014,Stellmach2014,DeWit2022}. There, the LSVs emerge as a result of the upscale kinetic energy transfer that feeds the largest available scales of the flow, owing to the quasi-2D nature of rotating turbulence. Indeed, both the quadrupolar vortex observed here and the LSVs are large scale ordered structures imposed on a turbulent background and are strongly vertically coherent (see also e.g. Ref.~\cite{DeWit2020}). Moreover, it resides purely in the horizontal manifold of the flow, as is also clear from the experimentally obtained power spectra in Madonia \textit{et al.} \cite{Madonia2023}. This raises the question what is at the root of the formation of the quadrupolar vortex as observed here: whether it is purely driven by the evolution of the wall mode and its interaction with the bulk, or whether the flow structure is a consequence of upscale energy transport. Therefore, we next resort to a case with a larger aspect ratio $\Gamma=0.72$, which is predicted to give rise to a wall mode with wavenumber $m=2$ according to Eq.~\eqref{eq:wavenumber}, in order to investigate whether or not this also gives rise to a quadrupolar vortex.

\subsection{Large aspect ratio: decaying jets}
In this wider cylinder $\Gamma=0.72$, in accordance with the prediction from Eq.~\eqref{eq:wavenumber}, we find a wall mode with wavenumber $m=2$, see Fig.~\ref{fig:moar_angle_time}. Consequently, since this yields four interfaces between hot rising sections and cold sinking sections of the wall mode, we observe four inward jets coming from the wall region, as opposed to the two jets observed in the smaller cylinder. These jets can best be appreciated from Video~\ref{vid:moar_jets_snap}.

\begin{video}[h!]
\href{https://youtu.be/nZ5FkRvjlCc}{\includegraphics[width=0.5\textwidth]{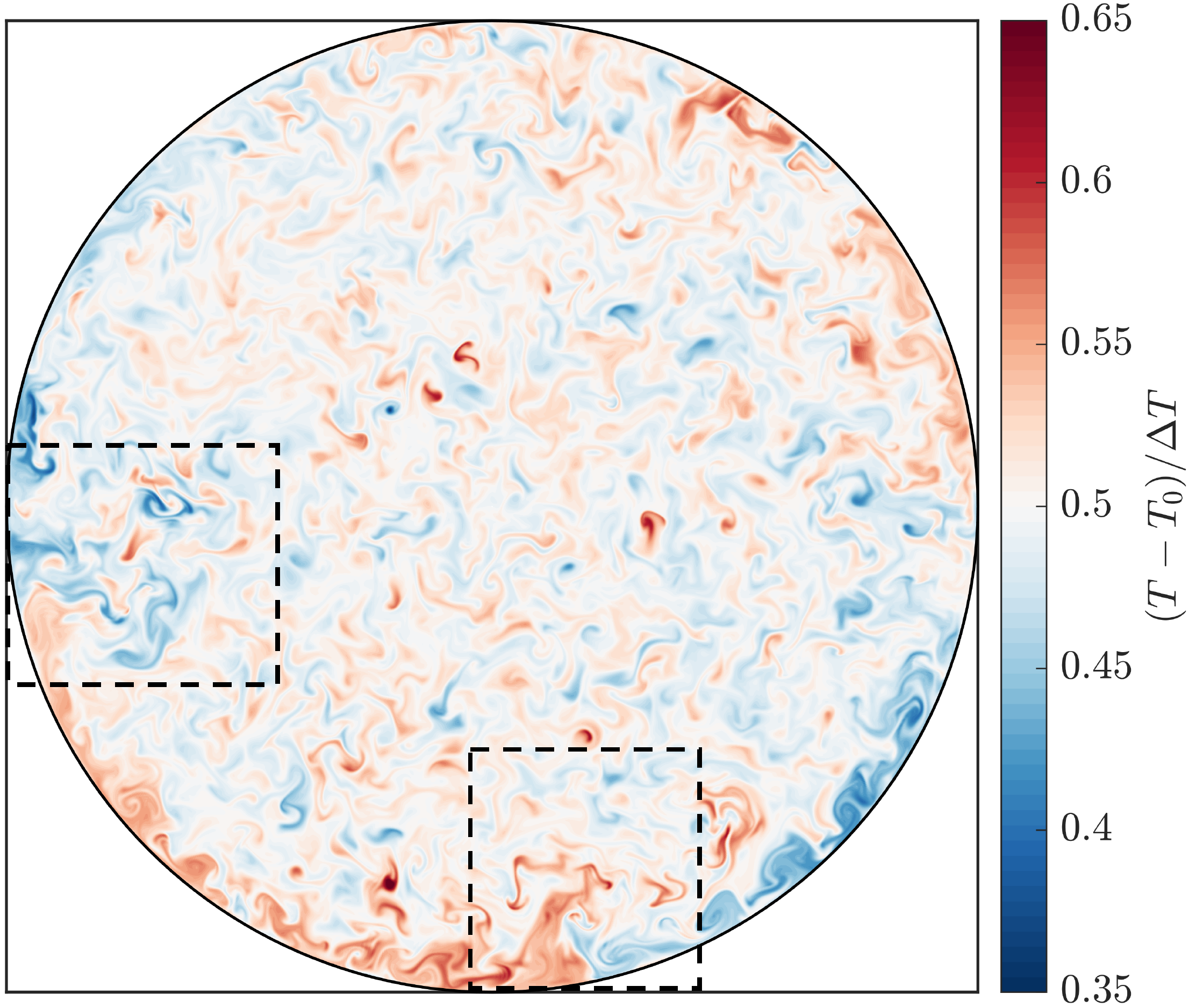}}%
 \setfloatlink{https://youtu.be/nZ5FkRvjlCc}%
 \caption{\label{vid:moar_jets_snap}Evolution of the temperature at mid-height for the the case with $\textrm{Ra}=2.0\times10^{11}$ and $\Gamma=0.72$, which shows the emergence of jets originating from the wall region that penetrate and decay into the bulk, highlighted by the dashed boxes.}
\end{video}

\begin{figure}[h!]
\includegraphics[width=1.0\textwidth]{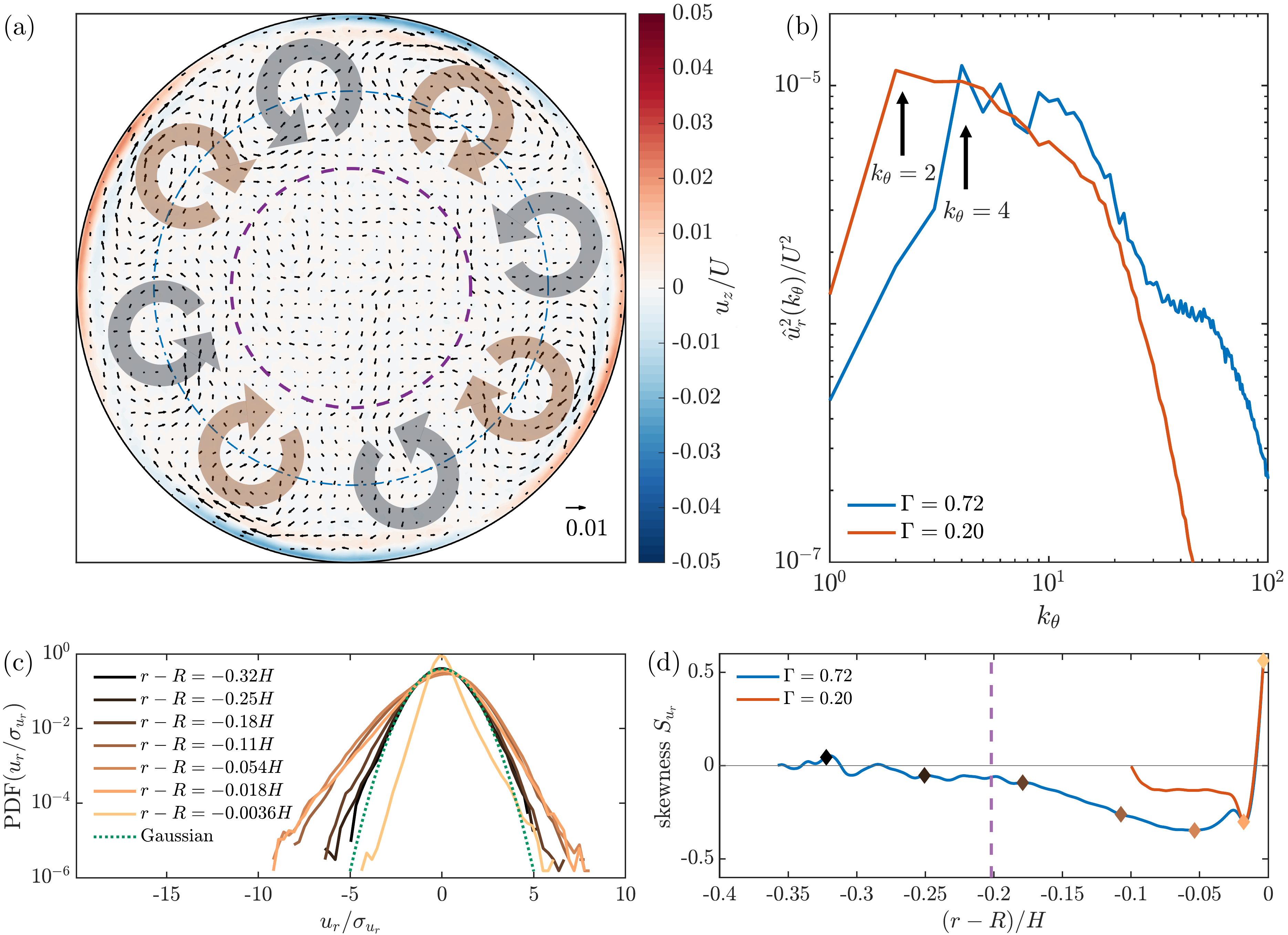}
\caption{\label{fig:jets_together}The orientation compensated mean flow (a) as well as azimuthal spectra (b), PDFs (c) and skewness (d) of radial velocity for the case with $\textrm{Ra}=2.0\times10^{11}$ and $\Gamma=0.72$ at mid-height. This depicts how jets emerging from the wall mode form a pattern in the azimuthal direction. The jets manifest as negative skewness in the radial velocity. They penetrate and decay into the bulk, thereby inducing an array of vortices in a region of width $W_J$ delineated by the purple dashed line in (a) and (d) as given by Eq.~\eqref{eq:jet_region}. The dashed-dotted blue line in (a) denotes the radius $R-W_J/2$ as used in Eq.~\eqref{eq:jet_region}. The azimuthal spectra in (b) are taken at $r=0.75R$. The dotted green line in (c) depicts the standard Gaussian distribution. Diamonds in (d) for the case $\Gamma=0.72$ indicate the corresponding radial positions for the PDFs in (c). In (b) and (d) the case with $\textrm{Ra}=2.1\times10^{11}$ and $\Gamma=0.20$ is also plotted for comparison.}
\end{figure}

Resorting to the orientation compensated average flow in Fig.~\ref{fig:jets_together}a and Appendix~\ref{app:moar_phase_ave}, we note that there is no formation of a quadrupolar vortex. Instead, in the larger cylinder, we observe an array of eight vortices with alternating sign of circulation in a region adjacent to the side wall, albeit less pronounced than the quadrupolar vortex in the smaller cylinder. This structure is also evident from the azimuthal energy spectrum $\hat{u}_r^2(k_\theta)$ of radial velocity as provided in Fig.~\ref{fig:jets_together}b, showing a peak around azimuthal wavenumber $k_\theta=4$ for the larger aspect ratio, while the smaller aspect ratio peaks around $k_\theta=2$, corresponding with, respectively, four inward jets for the 8-vortex and two inward jets for the quadrupolar vortex. These results suggest that the observed average flow is primarily due to the interaction of the wall mode with the bulk, rather than being driven by any upscale energy transfer.

This raises the following picture of how the bulk turbulence is affected by the jets emerging from the wall mode. In the smaller cylinder with $m=1$, jets emerging from the wall mode travel persistently to the center of the cylinder where they meet and recirculate into the quadrupolar vortex structure. In the wider cylinder with $m=2$, the jets penetrate into the bulk but decay before reaching the center, forming an opposite vortex on either side of the decaying jet, giving rise to the observed array of eight vortices.

The observation of the radial decay of the jets can be made quantitative by considering the skewness of the radial velocity $S_{u_r}=\langle [(u_r-\langle u_r\rangle)/\sigma_{u_r}]^3\rangle$ where $\sigma_{u_r}=\langle u_r^2-\langle u_r \rangle^2\rangle^{1/2}$ and $\langle...\rangle$ denotes temporal and azimuthal averaging. The relatively strong inward radial jets emerging from the wall mode manifest as negative skew in the PDF of radial velocity, see Fig.~\ref{fig:jets_together}c-d. We can estimate the width $W_J$ of the region affected by the decaying jets as the band that fits this array of $4m$ vortices, yielding
\begin{equation}\label{eq:jet_region}
\frac{2\pi(R-W_J/2)}{4m}\approx W_J \quad \Rightarrow \quad W_J\approx\frac{2\pi R}{4m + \pi} \sim \lambda_w,
\end{equation}
where we tacitly assumed that the vortices are as wide radially as azimuthally. Here, $\lambda_w$ denotes the wavelength $\lambda_w=2\pi R/m$ of the wall mode. The obtained $W_J$ is depicted in Fig.~\ref{fig:jets_together} as the dashed purple line, from which it is evident that it captures the knee in the radial velocity skewness well.

In general, we thus expect that the region that is affected by the penetrating jets of the wall mode scales as the wavelength of the wall mode. This implies that in order to obtain bulk turbulence that is largely unaffected by the wall mode, a high wavenumber of the wall mode is needed, or indeed, equivalently, a large aspect ratio of the convective flow domain. An alternative pathway to diminishing the effect of the wall mode was recently put forward in Ref.~\cite{Terrien2023}, suggesting to include narrow fins along the sidewall that suppress the development of the wall mode.

The final point we remark about the jet structure is the asymmetry of the temperature of the jets. As can be observed from Figs.~\ref{fig:small_phase_ave}d and \ref{fig:moar_phase_ave}d in Appendix~\ref{app:phase_ave}, jets that have the hot rising section of the wall mode on their left and the cold sinking section on their right (facing inward) are primarily hot jets and vice-versa for primarily cold jets. This suggests that the jets originate predominantly from the trailing edge of each section of the wall mode, relative to its anti-cyclonic precession direction.

\section{Reynolds decomposition: generation of zonal flow}\label{sec:RANS}
As the wall mode evolves from its linear onset into the turbulent regime under the influence of non-linear interactions, also the morphology of the wall mode itself changes. One peculiar aspect of the non-linear evolution of the wall mode is the development of a non-zero mean flow in the azimuthal direction as was emphasized in Ref.~\cite{Zhang2020} and also observed before in Refs.~\cite{Kunnen2011,Kunnen2013}. This non-zero mean azimuthal velocity in the wall region is observed when taking the ensemble average, even though this means that the average is taken over the entire phase (or azimuthal extent) of the wall mode. Refs.~\cite{Liao2006,Favier2020} put forward that Reynolds stresses can act as the source terms for this mean zonal flow. It is therefore instructive to consider the full Reynolds averaged Navier-Stokes (RANS) equations to understand how the non-linear interactions drive the mean flow in the wall mode.

We start by decomposing the flow into an ensemble averaged component $\boldsymbol{\mathrm{U}}$ and the perturbation  $\boldsymbol{u}'$ thereof, such that
\begin{equation}
    \boldsymbol{u}=\boldsymbol{\mathrm{U}}+\boldsymbol{u'}, \quad \langle \boldsymbol{u} \rangle = \boldsymbol{\mathrm{U}}, \quad \langle \boldsymbol{u}' \rangle=0.
\end{equation}
The RANS equations are then readily obtained by substituting this decomposition into the Navier-Stokes Eqs.~\eqref{eq:NS_momentum}~and~\eqref{eq:NS_cont} and taking the ensemble average. Due to the axial symmetry, any derivatives $\partial/\partial \theta$ vanish. We then find for the RANS $r$-equation (in dimensionless form, tildes are omitted for brevity)
\begin{subequations}
\begin{multline}\label{RANS_r}
    \frac{\partial \mathrm{U}_r}{\partial t} + \underbrace{\mathrm{U}_r\frac{\partial \mathrm{U}_r}{\partial r} + \mathrm{U}_z\frac{\partial \mathrm{U}_r}{\partial z} - \frac{\mathrm{U}_\theta^2}{r}}_{\textrm{advection}} + \underbrace{\frac{\partial}{\partial r}\left\langle u_r'^2 \right\rangle + \frac{\partial}{\partial z}\left\langle u_r' u_z'\right\rangle -\frac{\left\langle u_\theta'^2\right\rangle}{r}+\frac{\left\langle u_r'^2\right\rangle}{r}}_{\textrm{Reynolds stress}} + \underbrace{\frac{\partial\left\langle p\right\rangle}{\partial r}}_{\textrm{pressure}}\underbrace{-\frac{1}{\textrm{Ro}}\mathrm{U}_\theta}_{\textrm{Coriolis}} \\ + \underbrace{\sqrt{\frac{\textrm{Pr}}{\textrm{Ra}}}\left[-\frac{1}{r}\frac{\partial}{\partial r}\left(r\frac{\partial \mathrm{U}_r}{\partial r}\right) - \frac{\partial^2 \mathrm{U}_r}{\partial z^2} + \frac{\mathrm{U}_r}{r^2}\right]}_{\textrm{viscous}} = 0,
\end{multline}
and likewise for the $\theta$-equation
\begin{multline}\label{RANS_theta}
    \frac{\partial \mathrm{U}_\theta}{\partial t} + \underbrace{\mathrm{U}_r\frac{\partial \mathrm{U}_\theta}{\partial r} + \mathrm{U}_z\frac{\partial \mathrm{U}_\theta}{\partial z} + \frac{\mathrm{U}_r \textrm{ } \mathrm{U}_\theta}{r}}_{\textrm{advection}} + \underbrace{\frac{1}{r}\frac{\partial}{\partial r}\left(r \left\langle u_r' u_\theta'\right\rangle\right) + \frac{\partial}{\partial z}\left\langle u_\theta' u_z'\right\rangle + \frac{1}{r}\left\langle u_r' u_\theta'\right\rangle}_{\textrm{Reynolds stress}} + \underbrace{\frac{1}{\textrm{Ro}}\mathrm{U}_r}_{\textrm{Coriolis}} \\ + \underbrace{\sqrt{\frac{\textrm{Pr}}{\textrm{Ra}}}\left[-\frac{1}{r}\frac{\partial}{\partial r}\left(r\frac{\partial \mathrm{U}_\theta}{\partial r}\right) - \frac{\partial^2 \mathrm{U}_\theta}{\partial z^2} + \frac{\mathrm{U}_\theta}{r^2}\right]}_{\textrm{viscous}} = 0.
\end{multline}
\end{subequations}
In the statistically stationary state, the time derivatives $\partial/\partial t$ in the first terms vanish. Exploiting ergodicity, we approximate the ensemble average by a temporal and azimuthal average.

\begin{figure}[h!]
\includegraphics[width=0.95\textwidth]{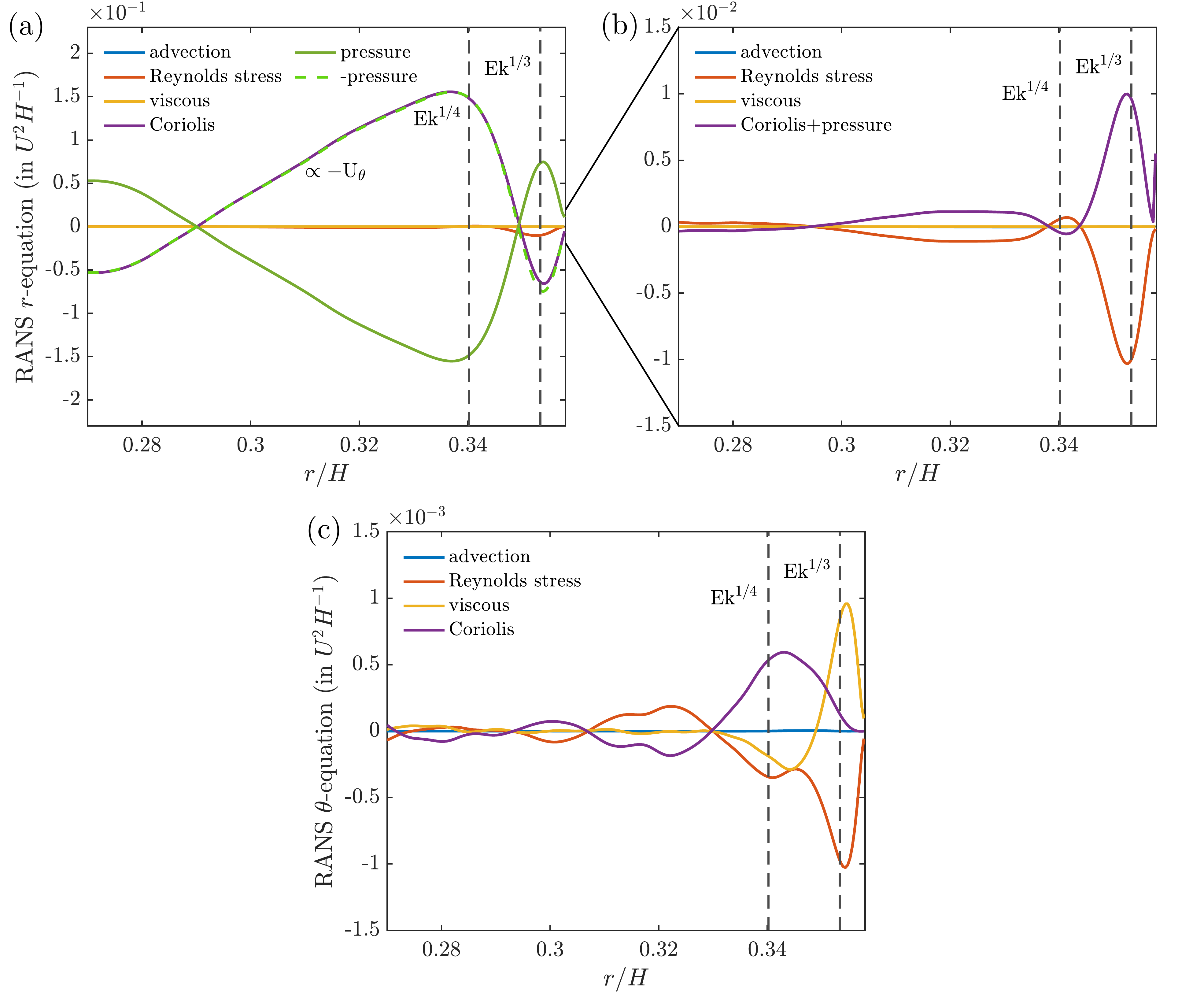}
\caption{\label{fig:moar_RANS}The different terms in the RANS $r$-equation (a,b) and $\theta$-equation (c) as given by Eqs.~\eqref{RANS_r}-\eqref{RANS_theta} for the case with $\textrm{Ra}=2.0\times10^{11}$ and $\Gamma=0.72$ at mid-height. The dashed vertical lines represent the Stewartson sidewall boundary layer thicknesses $\textrm{Ek}^{1/4}$ and $\textrm{Ek}^{1/3}$. The geostrophic balance in the $r$-equation between the Coriolis force and the pressure is shown in (a) as emphasized by the overlap between the Coriolis term and the negative of the pressure term (dashed green line), while the subleading balance is shown in the enlargement in (b) by combining the pressure and Coriolis terms. Note that the mean azimuthal flow $\mathrm{U}_\theta$ is directly proportional to the Coriolis term in (a).}
\end{figure}

The results for the different terms in the RANS equations in the wall region are provided in Fig.~\ref{fig:moar_RANS}. It reveals a two layer structure, approximately corresponding with the Stewartson boundary layer thicknesses of $\textrm{Ek}^{1/3}$ and $\textrm{Ek}^{1/4}$ \cite{Kunnen2011,Kunnen2013,Stewartson1957}. In the $r$-equation, as shown in Fig.~\ref{fig:moar_RANS}a, the leading order is the geostrophic balance $\partial \left\langle p\right\rangle /\partial r - (1/\textrm{Ro})\mathrm{U}_\theta = 0$ \cite{AguirreGuzman2021}. This is also the dominant balance that sustains the observed mean azimuthal flow, which is proportional to the Coriolis term in the $r$-equation. Considering the subleading balance in Fig.~\ref{fig:moar_RANS}b by combining the pressure and Coriolis terms, we find that while the inner layer (around $\textrm{Ek}^{1/4}$) is indeed primarily geostrophic, the outer layer (around $\textrm{Ek}^{1/3}$) has a more significant ageostrophic contribution that is balanced by Reynolds stresses in the $r$-equation. For the $\theta$-equation in Fig.~\ref{fig:moar_RANS}c, the force balance is yet an order of magnitude smaller. There, the balance in the outer layer is primarily between the Reynolds stresses and the viscous terms owing to gradients in the mean azimuthal flow, while the Coriolis force becomes important in the inner sidewall layer. The latter results from the slight net circulation of fluid being transported into the sidewall region near the top and bottom plates, which is transported radially inward at moderate heights, due to secondary circulation being set-up in the boundary layers \cite{Kunnen2013}. Reynolds stresses in the $\theta$-equation thus partly balance this radial flow through its Coriolis force and also drive the mean azimuthal flow through its viscous forces, while this azimuthal flow is largely amplified by geostrophic and ageostrophic forces in the $r$-equation.

We have considered this RANS balance for different simulation cases covered in this work and found qualitatively similar results for all considered $\textrm{Ra}$ and $\Gamma$ (not shown).

\section{Conclusions}\label{sec:conclusion}
In this work, we have numerically investigated the robust wall modes that exist in laterally confined rotating Rayleigh-B\'{e}nard convection. These strong zonal flow structures carry a large heat flux and precess anti-cyclonically along the sidewall of the convection tank. Understanding the characteristics of the wall mode and its interplay with the bulk rotating convective turbulence is important to the interpretation of experiments in the context of large scale geophysical and astrophysical convective flows.

We have analyzed the angular drift velocity of the wall mode and, by comparing our findings with earlier work, we have found a relation for the angular velocity that is entirely fit-parameter-free. It links the angular velocity of the wall mode to the critical angular velocity and $\textrm{Ra}$ at onset of the wall mode state, thereby establishing very strongly the surmised connection between the wall mode state that occurs before the onset of bulk convection and the boundary zonal flow in the turbulent regime: they are two sides of the same coin.

However, deviations from the obtained relation are also evident, showing that the drift velocity of the wall mode can decrease as the rotational constraint of the flow alleviates. We argue that this is a first manifestation of the break down of the wall mode, which is to be expected once the flow starts to move towards a buoyancy dominated regime and the stronger bulk turbulence starts to hamper the development of the wall mode. While this work provides first indications on where this break down can be expected, its exact functional dependence as well as its physical mechanism warrants further investigation.

Studying the interactive flow between the wall mode and the bulk turbulence, we find that this interaction is dominated by radial jets penetrating from the wall mode into the bulk. As these jets decay into the bulk turbulence, they induce opposite vortices on either side of the jet. For the cylinder with small aspect ratio, resulting in an $m=1$ wall mode, this results in the formation of a large quadrupolar vortex spanning the full domain. For the $m=2$ wall mode in the wider cylinder, this results in an array of eight (i.e. $4m$) alternating vortices adjacent to the sidewall, suggesting that the region affected by these penetrating jets scales as the wavelength of the wall mode.

The observations in the wider cylinder indicate that the vortex observed at the large scales in the bulk turbulence originate predominantly from the interaction with the wall mode, rather than being a result of an upscale flux of kinetic energy as is observed for unconfined rotating convection in a similar parameter range. Whether a stable upscale transport can be set up in the confined flow system remains to be explored in future work.

Finally, we consider the Reynolds averaged properties of the wall mode. There, we find that Reynolds stresses indeed play an essential role in the formation of the observed mean flow in the non-linear development of the wall mode. We find that the Reynolds stresses are largely balanced by Coriolis forces within the wall mode that dominate over viscous forces and advective forces, which would be more prominent in a conventional boundary layer.

The robust wall modes are a remarkable feature of confined rotating convection. This work has shown how properties of the wall mode at its onset propagate far into the turbulent regime for at least ten decades in $\textrm{Ra}$. We have explored the structure of the wall mode in this non-linear evolution and its interaction with the bulk turbulence. While we also identify pathways to mitigate the effect of the wall mode, indeed by resorting to a geometry that is wide compared to the wavelength of the wall mode, understanding the impact of the wall mode on the morphology of confined rotating convective turbulence will remain an important point of attention in current day and future research.

\begin{acknowledgments}
M.M., A.J.A.G. and R.P.J.K. received funding from the European Research Council (ERC) under the European Union’s Horizon 2020 research and innovation programme (Grant Agreement No. 678634). We are grateful for the support of the Netherlands Organisation for Scientific Research (NWO) for the use of supercomputer facilities (Cartesius and Snellius) under Grants No. 2019.005, No. 2020.009 and No. 2021.009. This publication is part of the project ``Shaping turbulence with smart particles'' with project number OCENW.GROOT.2019.031 of the research programme Open Competitie ENW XL which is (partly) financed by the Dutch Research Council (NWO).
\end{acknowledgments}

\clearpage
\appendix

\section{Comparison of scaling relation for wall mode frequency}\label{app:scaling_alt}

Here we compare the relation Eq.~\eqref{eq:ang_vel_scaling} for the angular precession frequency of the wall mode that is put forward in this work to the scaling relation $\omega_d/\Omega \propto \textrm{Ek}^{5/3} \textrm{Pr}^{-4/3}\textrm{Ra}$ that is obtained empirically by Zhang \textit{et al.} \cite{Zhang2021}. As argued in the main text, in the limit $\textrm{Ek}\to0$, relation Eq.~\eqref{eq:ang_vel_scaling} yields scaling exponents 2 and -1 respectively for $\textrm{Ek}$ and $\textrm{Pr}$, which differ only by a small amount from what is obtained by Zhang \textit{et al.} \cite{Zhang2021}. Since most data is obtained over a limited dynamic range in $\textrm{Ek}\sim\mathcal{O}(10^{-7}-10^{-6})$ and $\textrm{Pr}\sim\mathcal{O}(1)$, it is difficult to tell from the data alone which exponents capture the correct scaling.

\begin{figure}[h!]
\centering
\includegraphics[width=0.65\textwidth]{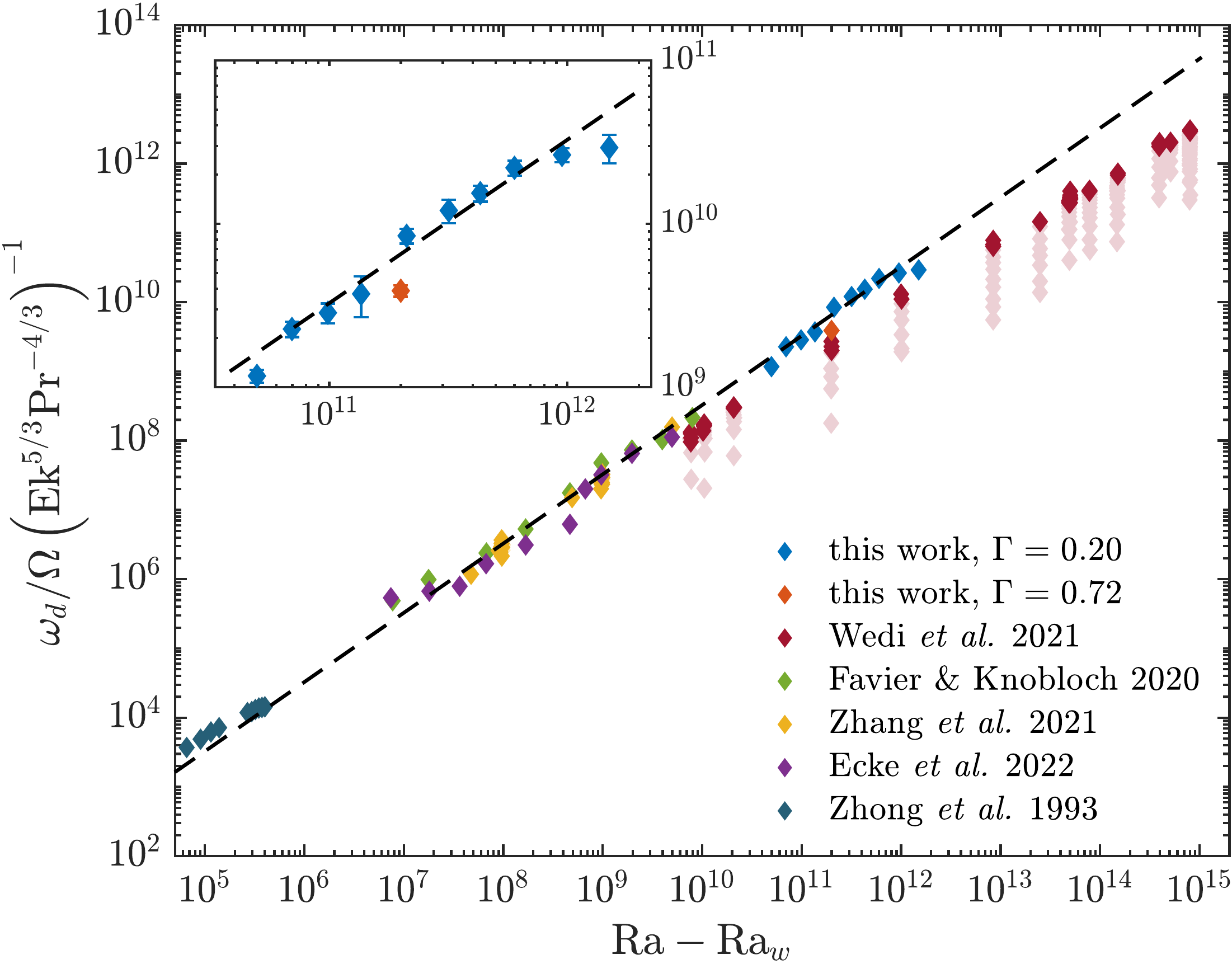}
\caption{\label{fig:ang_vel_all_alt}The angular drift velocity $\omega_d$ of the wall mode as a function of $\textrm{Ra}-\textrm{Ra}_w$ for our data and earlier works \cite{Zhang2021,Favier2020,Zhong1993,Ecke2022,Wedi2021}, compared with the scaling $\omega_d/\Omega \propto \textrm{Ek}^{5/3} \textrm{Pr}^{-4/3}\textrm{Ra}$ as put forward by Zhang \textit{et al.} \cite{Zhang2021} (dashed line). For the data from Wedi \textit{et al.} \cite{Wedi2021} and Zhang \textit{et al.} \cite{Zhang2021} data points with $\textrm{Ek}\textrm{Ra}^{1/2}>0.15$ are plotted in a lighter shade. The inset shows a close-up of our data including error bars. Compare with Fig.~\ref{fig:ang_vel_all} in the main text.}
\end{figure}

The scaling by Zhang \textit{et al.} \cite{Zhang2021} is shown in Fig.~\ref{fig:ang_vel_all_alt} for comparison with Fig.~\ref{fig:ang_vel_all} in the main text. Indeed, it is hard to say which one yields the better fit, although the data by Wedi \textit{et al.} \cite{Wedi2021} seems to be captured a bit better by the relation Eq.~\eqref{eq:ang_vel_scaling} in the main text of this work. However, we would argue that, based solely on the limited data available, one cannot tell which relation is correct.
In that light, we regard the comparison with the exponents obtained empirically by Zhang \textit{et al.} \cite{Zhang2021} as an approximate agreement, rather than an exact disagreement, given the limited data available at this stage.

\clearpage
\section{Orientation compensated average cross section}\label{app:phase_ave}

\begin{figure}[h!]
\includegraphics[width=0.8\textwidth]{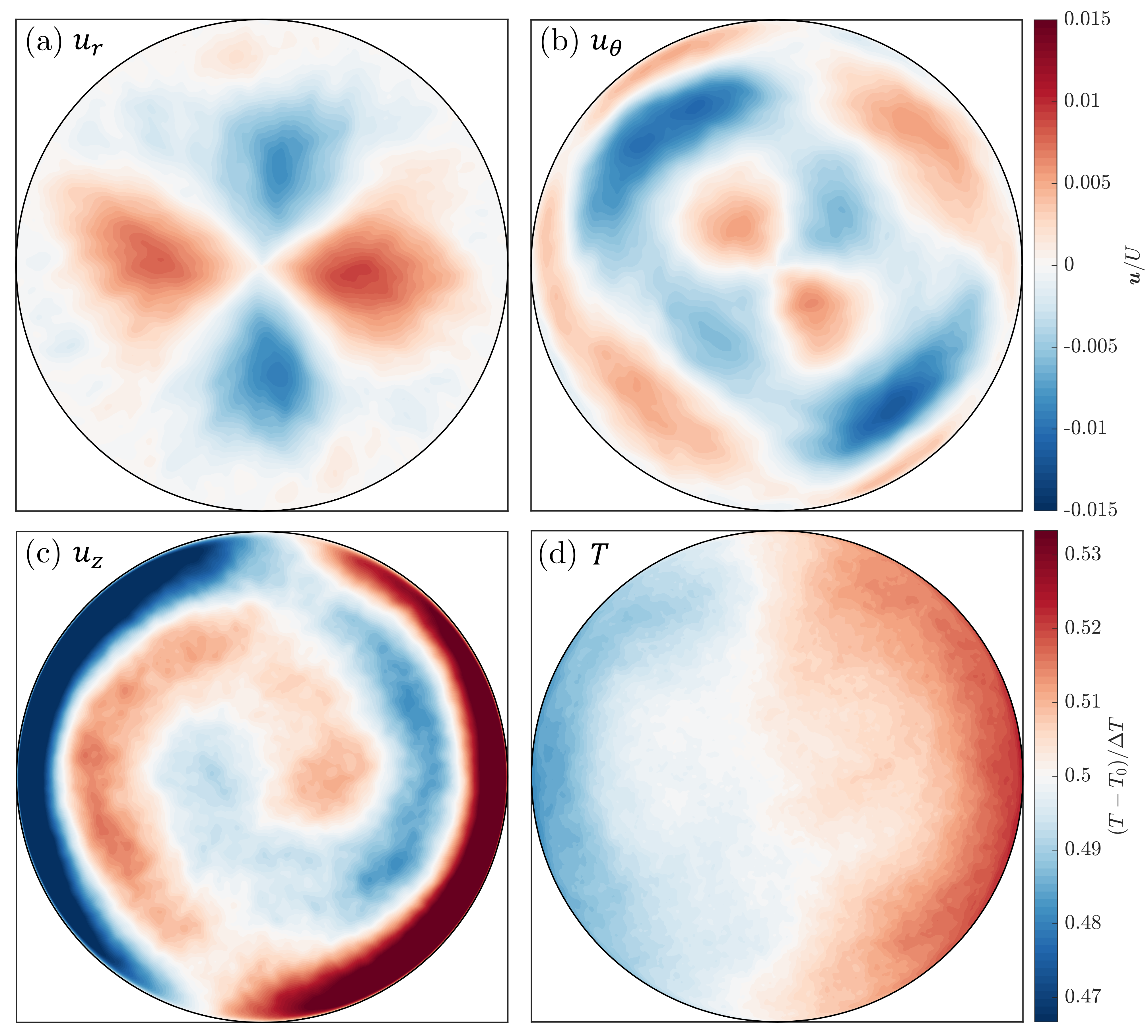}
\caption{\label{fig:small_phase_ave}Orientation compensated mean flow in the radial (a), azimuthal (b) and vertical (c) direction as well as the corresponding temperature (d) at mid-height for the case $\textrm{Ra}=9.5\times10^{11}$ and $\Gamma=0.20$.}
\end{figure}

\begin{figure}[h!]
\includegraphics[width=0.8\textwidth]{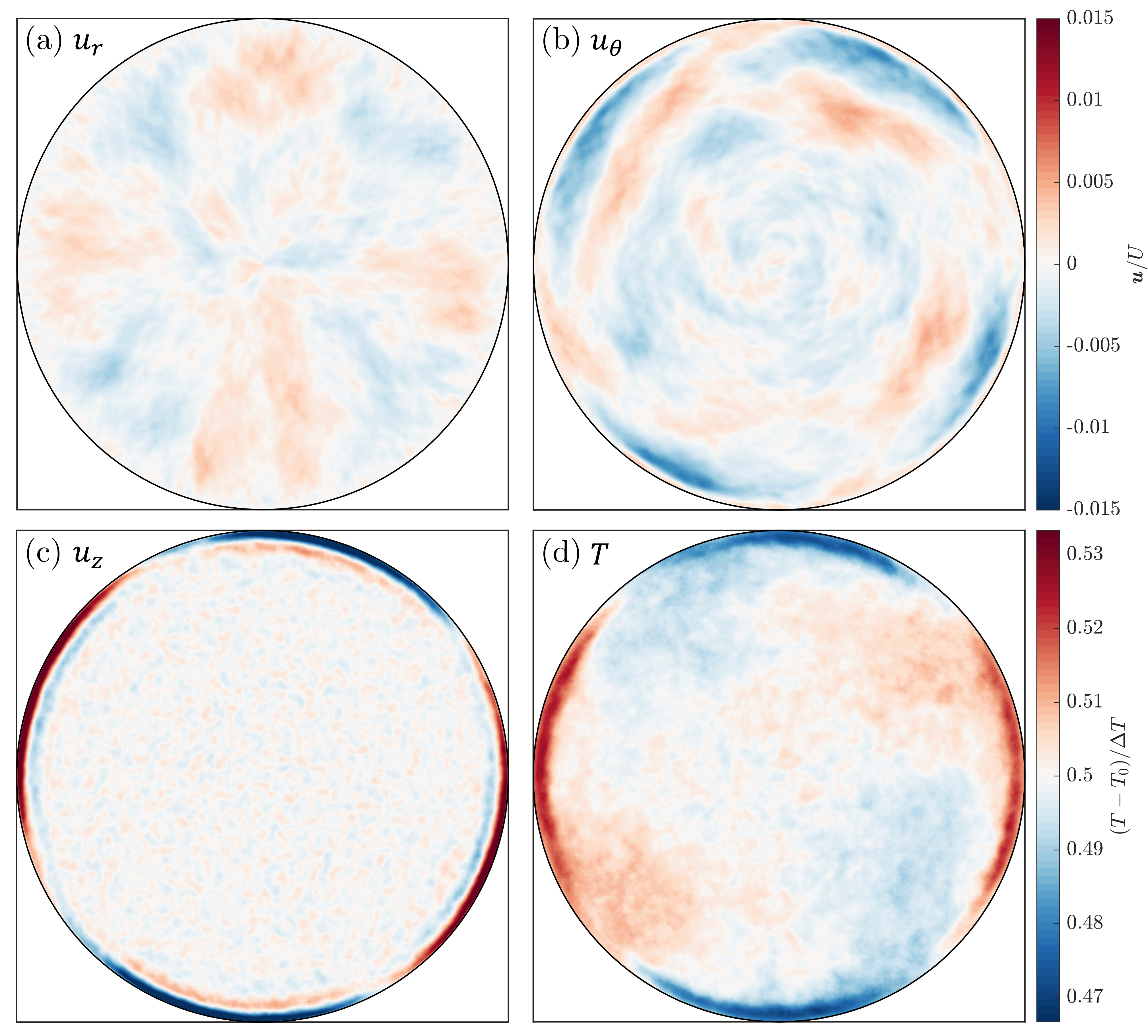}
\caption{\label{fig:moar_phase_ave}Orientation compensated mean flow in the radial (a), azimuthal (b) and vertical (c) direction as well as the corresponding temperature (d) at mid-height for the case $\textrm{Ra}=2.0\times10^{11}$ and $\Gamma=0.72$.}
\end{figure}

The different velocity components and temperature of the orientation compensated averages are provided for the small aspect ratio, Fig.~\ref{fig:small_phase_ave}, and large aspect ratio, Fig.~\ref{fig:moar_phase_ave}.

\subsection{Small aspect ratio}\label{app:small_phase_ave}

\clearpage
\subsection{Large aspect ratio}\label{app:moar_phase_ave}

\bibliography{main}

\begin{thebibliography}{41}%
\makeatletter
\providecommand \@ifxundefined [1]{%
 \@ifx{#1\undefined}
}%
\providecommand \@ifnum [1]{%
 \ifnum #1\expandafter \@firstoftwo
 \else \expandafter \@secondoftwo
 \fi
}%
\providecommand \@ifx [1]{%
 \ifx #1\expandafter \@firstoftwo
 \else \expandafter \@secondoftwo
 \fi
}%
\providecommand \natexlab [1]{#1}%
\providecommand \enquote  [1]{``#1''}%
\providecommand \bibnamefont  [1]{#1}%
\providecommand \bibfnamefont [1]{#1}%
\providecommand \citenamefont [1]{#1}%
\providecommand \href@noop [0]{\@secondoftwo}%
\providecommand \href [0]{\begingroup \@sanitize@url \@href}%
\providecommand \@href[1]{\@@startlink{#1}\@@href}%
\providecommand \@@href[1]{\endgroup#1\@@endlink}%
\providecommand \@sanitize@url [0]{\catcode `\\12\catcode `\$12\catcode
  `\&12\catcode `\#12\catcode `\^12\catcode `\_12\catcode `\%12\relax}%
\providecommand \@@startlink[1]{}%
\providecommand \@@endlink[0]{}%
\providecommand \url  [0]{\begingroup\@sanitize@url \@url }%
\providecommand \@url [1]{\endgroup\@href {#1}{\urlprefix }}%
\providecommand \urlprefix  [0]{URL }%
\providecommand \Eprint [0]{\href }%
\providecommand \doibase [0]{https://doi.org/}%
\providecommand \selectlanguage [0]{\@gobble}%
\providecommand \bibinfo  [0]{\@secondoftwo}%
\providecommand \bibfield  [0]{\@secondoftwo}%
\providecommand \translation [1]{[#1]}%
\providecommand \BibitemOpen [0]{}%
\providecommand \bibitemStop [0]{}%
\providecommand \bibitemNoStop [0]{.\EOS\space}%
\providecommand \EOS [0]{\spacefactor3000\relax}%
\providecommand \BibitemShut  [1]{\csname bibitem#1\endcsname}%
\let\auto@bib@innerbib\@empty
\bibitem [{\citenamefont {Aurnou}\ \emph {et~al.}(2015)\citenamefont {Aurnou},
  \citenamefont {Calkins}, \citenamefont {Cheng}, \citenamefont {Julien},
  \citenamefont {King}, \citenamefont {Nieves}, \citenamefont {Soderlund},\
  and\ \citenamefont {Stellmach}}]{Aurnou2015}%
  \BibitemOpen
  \bibfield  {author} {\bibinfo {author} {\bibfnamefont {J.~M.}\ \bibnamefont
  {Aurnou}}, \bibinfo {author} {\bibfnamefont {M.~A.}\ \bibnamefont {Calkins}},
  \bibinfo {author} {\bibfnamefont {J.~S.}\ \bibnamefont {Cheng}}, \bibinfo
  {author} {\bibfnamefont {K.}~\bibnamefont {Julien}}, \bibinfo {author}
  {\bibfnamefont {E.~M.}\ \bibnamefont {King}}, \bibinfo {author}
  {\bibfnamefont {D.}~\bibnamefont {Nieves}}, \bibinfo {author} {\bibfnamefont
  {K.~M.}\ \bibnamefont {Soderlund}},\ and\ \bibinfo {author} {\bibfnamefont
  {S.}~\bibnamefont {Stellmach}},\ }\bibfield  {title} {\bibinfo {title}
  {Rotating convective turbulence in earth and planetary cores},\ }\href
  {https://doi.org/https://doi.org/10.1016/j.pepi.2015.07.001} {\bibfield
  {journal} {\bibinfo  {journal} {Phys. Earth Planet. Inter.}\ }\textbf
  {\bibinfo {volume} {246}},\ \bibinfo {pages} {52\textendash71} (\bibinfo
  {year} {2015})}\BibitemShut {NoStop}%
\bibitem [{\citenamefont {Miesch}(2000)}]{Miesch2000}%
  \BibitemOpen
  \bibfield  {author} {\bibinfo {author} {\bibfnamefont {M.~S.}\ \bibnamefont
  {Miesch}},\ }\bibfield  {title} {\bibinfo {title} {{The coupling of solar
  convection and rotation}},\ }\href {https://doi.org/10.1023/a:1005260527450}
  {\bibfield  {journal} {\bibinfo  {journal} {Sol. Phys.}\ }\textbf {\bibinfo
  {volume} {192}},\ \bibinfo {pages} {59\textendash89} (\bibinfo {year}
  {2000})}\BibitemShut {NoStop}%
\bibitem [{\citenamefont {Heimpel}\ \emph {et~al.}(2005)\citenamefont
  {Heimpel}, \citenamefont {Aurnou},\ and\ \citenamefont
  {Wicht}}]{Heimpel2005}%
  \BibitemOpen
  \bibfield  {author} {\bibinfo {author} {\bibfnamefont {M.}~\bibnamefont
  {Heimpel}}, \bibinfo {author} {\bibfnamefont {J.}~\bibnamefont {Aurnou}},\
  and\ \bibinfo {author} {\bibfnamefont {J.}~\bibnamefont {Wicht}},\ }\bibfield
   {title} {\bibinfo {title} {{Simulation of equatorial and high-latitude jets
  on Jupiter in a deep convection model}},\ }\href
  {https://doi.org/10.1038/nature04208} {\bibfield  {journal} {\bibinfo
  {journal} {Nature}\ }\textbf {\bibinfo {volume} {438}},\ \bibinfo {pages}
  {193\textendash196} (\bibinfo {year} {2005})}\BibitemShut {NoStop}%
\bibitem [{\citenamefont {Ecke}\ and\ \citenamefont
  {Shishkina}(2023)}]{Ecke2023}%
  \BibitemOpen
  \bibfield  {author} {\bibinfo {author} {\bibfnamefont {R.~E.}\ \bibnamefont
  {Ecke}}\ and\ \bibinfo {author} {\bibfnamefont {O.}~\bibnamefont
  {Shishkina}},\ }\bibfield  {title} {\bibinfo {title} {Turbulent rotating
  rayleigh–bénard convection},\ }\href
  {https://doi.org/10.1146/annurev-fluid-120720-020446} {\bibfield  {journal}
  {\bibinfo  {journal} {Annu. Rev. Fluid Mech.}\ }\textbf {\bibinfo {volume}
  {55}},\ \bibinfo {pages} {603\textendash638} (\bibinfo {year}
  {2023})}\BibitemShut {NoStop}%
\bibitem [{\citenamefont {Kunnen}(2021)}]{Kunnen2021}%
  \BibitemOpen
  \bibfield  {author} {\bibinfo {author} {\bibfnamefont {R.~P.~J.}\
  \bibnamefont {Kunnen}},\ }\bibfield  {title} {\bibinfo {title} {{The
  geostrophic regime of rapidly rotating turbulent convection}},\ }\href
  {https://doi.org/10.1080/14685248.2021.1876877} {\bibfield  {journal}
  {\bibinfo  {journal} {J. Turbulence}\ }\textbf {\bibinfo {volume} {22}},\
  \bibinfo {pages} {267\textendash296} (\bibinfo {year} {2021})}\BibitemShut
  {NoStop}%
\bibitem [{\citenamefont {Bouillaut}\ \emph {et~al.}(2021)\citenamefont
  {Bouillaut}, \citenamefont {Miquel}, \citenamefont {Julien}, \citenamefont
  {Auma{\^{i}}tre},\ and\ \citenamefont {Gallet}}]{Bouillaut2021}%
  \BibitemOpen
  \bibfield  {author} {\bibinfo {author} {\bibfnamefont {V.}~\bibnamefont
  {Bouillaut}}, \bibinfo {author} {\bibfnamefont {B.}~\bibnamefont {Miquel}},
  \bibinfo {author} {\bibfnamefont {K.}~\bibnamefont {Julien}}, \bibinfo
  {author} {\bibfnamefont {S.}~\bibnamefont {Auma{\^{i}}tre}},\ and\ \bibinfo
  {author} {\bibfnamefont {B.}~\bibnamefont {Gallet}},\ }\bibfield  {title}
  {\bibinfo {title} {Experimental observation of the geostrophic turbulence
  regime of rapidly rotating convection},\ }\href
  {https://doi.org/10.1073/pnas.2105015118} {\bibfield  {journal} {\bibinfo
  {journal} {Proc. Natl. Acad. Sci.}\ }\textbf {\bibinfo {volume} {118}},\
  \bibinfo {pages} {e2105015118} (\bibinfo {year} {2021})}\BibitemShut
  {NoStop}%
\bibitem [{\citenamefont {Madonia}\ \emph {et~al.}(2021)\citenamefont
  {Madonia}, \citenamefont {{Aguirre Guzm{\'{a}}n}}, \citenamefont {Clercx},\
  and\ \citenamefont {Kunnen}}]{Madonia2021}%
  \BibitemOpen
  \bibfield  {author} {\bibinfo {author} {\bibfnamefont {M.}~\bibnamefont
  {Madonia}}, \bibinfo {author} {\bibfnamefont {A.~J.}\ \bibnamefont {{Aguirre
  Guzm{\'{a}}n}}}, \bibinfo {author} {\bibfnamefont {H.~J.~H.}\ \bibnamefont
  {Clercx}},\ and\ \bibinfo {author} {\bibfnamefont {R.~P.~J.}\ \bibnamefont
  {Kunnen}},\ }\bibfield  {title} {\bibinfo {title} {{Velocimetry in rapidly
  rotating convection: Spatial correlations, flow structures and length
  scales}},\ }\href {https://doi.org/10.1209/0295-5075/AC30D6} {\bibfield
  {journal} {\bibinfo  {journal} {Europhys. Lett.}\ }\textbf {\bibinfo {volume}
  {135}},\ \bibinfo {pages} {54002} (\bibinfo {year} {2021})}\BibitemShut
  {NoStop}%
\bibitem [{\citenamefont {Wedi}\ \emph {et~al.}(2021)\citenamefont {Wedi},
  \citenamefont {van Gils}, \citenamefont {Bodenschatz},\ and\ \citenamefont
  {Weiss}}]{Wedi2021}%
  \BibitemOpen
  \bibfield  {author} {\bibinfo {author} {\bibfnamefont {M.}~\bibnamefont
  {Wedi}}, \bibinfo {author} {\bibfnamefont {D.~P.~M.}\ \bibnamefont {van
  Gils}}, \bibinfo {author} {\bibfnamefont {E.}~\bibnamefont {Bodenschatz}},\
  and\ \bibinfo {author} {\bibfnamefont {S.}~\bibnamefont {Weiss}},\ }\bibfield
   {title} {\bibinfo {title} {{Rotating turbulent thermal convection at very
  large Rayleigh numbers}},\ }\href {https://doi.org/10.1017/JFM.2020.1149}
  {\bibfield  {journal} {\bibinfo  {journal} {J. Fluid Mech.}\ }\textbf
  {\bibinfo {volume} {912}},\ \bibinfo {pages} {A30} (\bibinfo {year}
  {2021})}\BibitemShut {NoStop}%
\bibitem [{\citenamefont {Stellmach}\ \emph {et~al.}(2014)\citenamefont
  {Stellmach}, \citenamefont {Lischper}, \citenamefont {Julien}, \citenamefont
  {Vasil}, \citenamefont {Cheng}, \citenamefont {Ribeiro}, \citenamefont
  {King},\ and\ \citenamefont {Aurnou}}]{Stellmach2014}%
  \BibitemOpen
  \bibfield  {author} {\bibinfo {author} {\bibfnamefont {S.}~\bibnamefont
  {Stellmach}}, \bibinfo {author} {\bibfnamefont {M.}~\bibnamefont {Lischper}},
  \bibinfo {author} {\bibfnamefont {K.}~\bibnamefont {Julien}}, \bibinfo
  {author} {\bibfnamefont {G.}~\bibnamefont {Vasil}}, \bibinfo {author}
  {\bibfnamefont {J.~S.}\ \bibnamefont {Cheng}}, \bibinfo {author}
  {\bibfnamefont {A.}~\bibnamefont {Ribeiro}}, \bibinfo {author} {\bibfnamefont
  {E.~M.}\ \bibnamefont {King}},\ and\ \bibinfo {author} {\bibfnamefont
  {J.~M.}\ \bibnamefont {Aurnou}},\ }\bibfield  {title} {\bibinfo {title}
  {{Approaching the asymptotic regime of rapidly rotating convection: Boundary
  layers versus interior dynamics}},\ }\href
  {https://doi.org/10.1103/PhysRevLett.113.254501} {\bibfield  {journal}
  {\bibinfo  {journal} {Phys. Rev. Lett.}\ }\textbf {\bibinfo {volume} {113}},\
  \bibinfo {pages} {254501} (\bibinfo {year} {2014})}\BibitemShut {NoStop}%
\bibitem [{\citenamefont {Maffei}\ \emph {et~al.}(2021)\citenamefont {Maffei},
  \citenamefont {Krouss}, \citenamefont {Julien},\ and\ \citenamefont
  {Calkins}}]{Maffei2021}%
  \BibitemOpen
  \bibfield  {author} {\bibinfo {author} {\bibfnamefont {S.}~\bibnamefont
  {Maffei}}, \bibinfo {author} {\bibfnamefont {M.~J.}\ \bibnamefont {Krouss}},
  \bibinfo {author} {\bibfnamefont {K.}~\bibnamefont {Julien}},\ and\ \bibinfo
  {author} {\bibfnamefont {M.~A.}\ \bibnamefont {Calkins}},\ }\bibfield
  {title} {\bibinfo {title} {{On the inverse cascade and flow speed scaling
  behavior in rapidly rotating Rayleigh-B{\'{e}}nard convection}},\ }\href
  {https://doi.org/10.1017/jfm.2020.1058} {\bibfield  {journal} {\bibinfo
  {journal} {J. Fluid Mech.}\ }\textbf {\bibinfo {volume} {913}},\ \bibinfo
  {pages} {A18} (\bibinfo {year} {2021})}\BibitemShut {NoStop}%
\bibitem [{\citenamefont {Aguirre~Guzm\'an}\ \emph {et~al.}(2022)\citenamefont
  {Aguirre~Guzm\'an}, \citenamefont {Madonia}, \citenamefont {Cheng},
  \citenamefont {Ostilla-M\'onico}, \citenamefont {Clercx},\ and\ \citenamefont
  {Kunnen}}]{AguirreGuzman2022}%
  \BibitemOpen
  \bibfield  {author} {\bibinfo {author} {\bibfnamefont {A.~J.}\ \bibnamefont
  {Aguirre~Guzm\'an}}, \bibinfo {author} {\bibfnamefont {M.}~\bibnamefont
  {Madonia}}, \bibinfo {author} {\bibfnamefont {J.~S.}\ \bibnamefont {Cheng}},
  \bibinfo {author} {\bibfnamefont {R.}~\bibnamefont {Ostilla-M\'onico}},
  \bibinfo {author} {\bibfnamefont {H.~J.~H.}\ \bibnamefont {Clercx}},\ and\
  \bibinfo {author} {\bibfnamefont {R.~P.~J.}\ \bibnamefont {Kunnen}},\
  }\bibfield  {title} {\bibinfo {title} {Flow- and temperature-based statistics
  characterizing the regimes in rapidly rotating turbulent convection in
  simulations employing no-slip boundary conditions},\ }\href
  {https://doi.org/10.1103/PhysRevFluids.7.013501} {\bibfield  {journal}
  {\bibinfo  {journal} {Phys. Rev. Fluids}\ }\textbf {\bibinfo {volume} {7}},\
  \bibinfo {pages} {013501} (\bibinfo {year} {2022})}\BibitemShut {NoStop}%
\bibitem [{\citenamefont {de~Wit}\ \emph {et~al.}(2020)\citenamefont {de~Wit},
  \citenamefont {Aguirre~Guzm{\'{a}}n}, \citenamefont {Madonia}, \citenamefont
  {Cheng}, \citenamefont {Clercx},\ and\ \citenamefont {Kunnen}}]{DeWit2020}%
  \BibitemOpen
  \bibfield  {author} {\bibinfo {author} {\bibfnamefont {X.~M.}\ \bibnamefont
  {de~Wit}}, \bibinfo {author} {\bibfnamefont {A.~J.}\ \bibnamefont
  {Aguirre~Guzm{\'{a}}n}}, \bibinfo {author} {\bibfnamefont {M.}~\bibnamefont
  {Madonia}}, \bibinfo {author} {\bibfnamefont {J.~S.}\ \bibnamefont {Cheng}},
  \bibinfo {author} {\bibfnamefont {H.~J.~H.}\ \bibnamefont {Clercx}},\ and\
  \bibinfo {author} {\bibfnamefont {R.~P.~J.}\ \bibnamefont {Kunnen}},\
  }\bibfield  {title} {\bibinfo {title} {{Turbulent rotating convection
  confined in a slender cylinder: the sidewall circulation}},\ }\href
  {https://doi.org/10.1103/PhysRevFluids.5.023502} {\bibfield  {journal}
  {\bibinfo  {journal} {Phys. Rev. Fluids}\ }\textbf {\bibinfo {volume} {5}},\
  \bibinfo {pages} {023502} (\bibinfo {year} {2020})}\BibitemShut {NoStop}%
\bibitem [{\citenamefont {Zhang}\ \emph {et~al.}(2020)\citenamefont {Zhang},
  \citenamefont {van Gils}, \citenamefont {Horn}, \citenamefont {Wedi},
  \citenamefont {Zwirner}, \citenamefont {Ahlers}, \citenamefont {Ecke},
  \citenamefont {Weiss}, \citenamefont {Bodenschatz},\ and\ \citenamefont
  {Shishkina}}]{Zhang2020}%
  \BibitemOpen
  \bibfield  {author} {\bibinfo {author} {\bibfnamefont {X.}~\bibnamefont
  {Zhang}}, \bibinfo {author} {\bibfnamefont {D.~P.~M.}\ \bibnamefont {van
  Gils}}, \bibinfo {author} {\bibfnamefont {S.}~\bibnamefont {Horn}}, \bibinfo
  {author} {\bibfnamefont {M.}~\bibnamefont {Wedi}}, \bibinfo {author}
  {\bibfnamefont {L.}~\bibnamefont {Zwirner}}, \bibinfo {author} {\bibfnamefont
  {G.}~\bibnamefont {Ahlers}}, \bibinfo {author} {\bibfnamefont {R.~E.}\
  \bibnamefont {Ecke}}, \bibinfo {author} {\bibfnamefont {S.}~\bibnamefont
  {Weiss}}, \bibinfo {author} {\bibfnamefont {E.}~\bibnamefont {Bodenschatz}},\
  and\ \bibinfo {author} {\bibfnamefont {O.}~\bibnamefont {Shishkina}},\
  }\bibfield  {title} {\bibinfo {title} {{Boundary Zonal Flow in Rotating
  Turbulent Rayleigh-B{\'{e}}nard Convection}},\ }\href
  {https://doi.org/10.1103/PhysRevLett.124.084505} {\bibfield  {journal}
  {\bibinfo  {journal} {Phys. Rev. Lett.}\ }\textbf {\bibinfo {volume} {124}},\
  \bibinfo {pages} {84505} (\bibinfo {year} {2020})}\BibitemShut {NoStop}%
\bibitem [{\citenamefont {Favier}\ and\ \citenamefont
  {Knobloch}(2020)}]{Favier2020}%
  \BibitemOpen
  \bibfield  {author} {\bibinfo {author} {\bibfnamefont {B.}~\bibnamefont
  {Favier}}\ and\ \bibinfo {author} {\bibfnamefont {E.}~\bibnamefont
  {Knobloch}},\ }\bibfield  {title} {\bibinfo {title} {{Robust wall states in
  rapidly rotating Rayleigh-B{\'{e}}nard convection}},\ }\href
  {https://doi.org/10.1017/jfm.2020.310} {\bibfield  {journal} {\bibinfo
  {journal} {J. Fluid Mech.}\ }\textbf {\bibinfo {volume} {895}},\ \bibinfo
  {pages} {R1} (\bibinfo {year} {2020})}\BibitemShut {NoStop}%
\bibitem [{\citenamefont {Shishkina}(2020)}]{Shishkina2020}%
  \BibitemOpen
  \bibfield  {author} {\bibinfo {author} {\bibfnamefont {O.}~\bibnamefont
  {Shishkina}},\ }\bibfield  {title} {\bibinfo {title} {{Tenacious wall states
  in thermal convection in rapidly rotating containers}},\ }\href
  {https://doi.org/10.1017/JFM.2020.420} {\bibfield  {journal} {\bibinfo
  {journal} {J. Fluid Mech.}\ }\textbf {\bibinfo {volume} {898}},\ \bibinfo
  {pages} {F1} (\bibinfo {year} {2020})}\BibitemShut {NoStop}%
\bibitem [{\citenamefont {Zhang}\ \emph {et~al.}(2021)\citenamefont {Zhang},
  \citenamefont {Ecke},\ and\ \citenamefont {Shishkina}}]{Zhang2021}%
  \BibitemOpen
  \bibfield  {author} {\bibinfo {author} {\bibfnamefont {X.}~\bibnamefont
  {Zhang}}, \bibinfo {author} {\bibfnamefont {R.~E.}\ \bibnamefont {Ecke}},\
  and\ \bibinfo {author} {\bibfnamefont {O.}~\bibnamefont {Shishkina}},\
  }\bibfield  {title} {\bibinfo {title} {{Boundary zonal flows in rapidly
  rotating turbulent thermal convection}},\ }\href
  {https://doi.org/10.1017/JFM.2021.74} {\bibfield  {journal} {\bibinfo
  {journal} {J. Fluid Mech.}\ }\textbf {\bibinfo {volume} {915}},\ \bibinfo
  {pages} {A62} (\bibinfo {year} {2021})}\BibitemShut {NoStop}%
\bibitem [{\citenamefont {Lu}\ \emph {et~al.}(2021)\citenamefont {Lu},
  \citenamefont {Ding}, \citenamefont {Shi}, \citenamefont {Xia},\ and\
  \citenamefont {Zhong}}]{Lu2021}%
  \BibitemOpen
  \bibfield  {author} {\bibinfo {author} {\bibfnamefont {H.-Y.}\ \bibnamefont
  {Lu}}, \bibinfo {author} {\bibfnamefont {G.-Y.}\ \bibnamefont {Ding}},
  \bibinfo {author} {\bibfnamefont {J.-Q.}\ \bibnamefont {Shi}}, \bibinfo
  {author} {\bibfnamefont {K.-Q.}\ \bibnamefont {Xia}},\ and\ \bibinfo {author}
  {\bibfnamefont {J.-Q.}\ \bibnamefont {Zhong}},\ }\bibfield  {title} {\bibinfo
  {title} {{Heat-transport scaling and transition in geostrophic rotating
  convection with varying aspect ratio}},\ }\href
  {https://doi.org/10.1103/PhysRevFluids.6.L071501} {\bibfield  {journal}
  {\bibinfo  {journal} {Phys. Rev. Fluids}\ }\textbf {\bibinfo {volume} {6}},\
  \bibinfo {pages} {L071501} (\bibinfo {year} {2021})}\BibitemShut {NoStop}%
\bibitem [{\citenamefont {Wedi}\ \emph {et~al.}(2022)\citenamefont {Wedi},
  \citenamefont {Moturi}, \citenamefont {Funfschilling},\ and\ \citenamefont
  {Weiss}}]{Wedi2022}%
  \BibitemOpen
  \bibfield  {author} {\bibinfo {author} {\bibfnamefont {M.}~\bibnamefont
  {Wedi}}, \bibinfo {author} {\bibfnamefont {V.~M.}\ \bibnamefont {Moturi}},
  \bibinfo {author} {\bibfnamefont {D.}~\bibnamefont {Funfschilling}},\ and\
  \bibinfo {author} {\bibfnamefont {S.}~\bibnamefont {Weiss}},\ }\bibfield
  {title} {\bibinfo {title} {{Experimental evidence for the boundary zonal flow
  in rotating Rayleigh–B{\'{e}}nard convection}},\ }\href
  {https://doi.org/10.1017/JFM.2022.195} {\bibfield  {journal} {\bibinfo
  {journal} {J. Fluid Mech.}\ }\textbf {\bibinfo {volume} {939}},\ \bibinfo
  {pages} {A14} (\bibinfo {year} {2022})}\BibitemShut {NoStop}%
\bibitem [{\citenamefont {Terrien}\ \emph {et~al.}(2023)\citenamefont
  {Terrien}, \citenamefont {Favier},\ and\ \citenamefont
  {Knobloch}}]{Terrien2023}%
  \BibitemOpen
  \bibfield  {author} {\bibinfo {author} {\bibfnamefont {L.}~\bibnamefont
  {Terrien}}, \bibinfo {author} {\bibfnamefont {B.}~\bibnamefont {Favier}},\
  and\ \bibinfo {author} {\bibfnamefont {E.}~\bibnamefont {Knobloch}},\
  }\bibfield  {title} {\bibinfo {title} {Suppression of wall modes in rapidly
  rotating rayleigh-b\'enard convection by narrow horizontal fins},\ }\href
  {https://doi.org/10.1103/PhysRevLett.130.174002} {\bibfield  {journal}
  {\bibinfo  {journal} {Phys. Rev. Lett.}\ }\textbf {\bibinfo {volume} {130}},\
  \bibinfo {pages} {174002} (\bibinfo {year} {2023})}\BibitemShut {NoStop}%
\bibitem [{\citenamefont {Ecke}\ \emph {et~al.}(2022)\citenamefont {Ecke},
  \citenamefont {Zhang},\ and\ \citenamefont {Shishkina}}]{Ecke2022}%
  \BibitemOpen
  \bibfield  {author} {\bibinfo {author} {\bibfnamefont {R.~E.}\ \bibnamefont
  {Ecke}}, \bibinfo {author} {\bibfnamefont {X.}~\bibnamefont {Zhang}},\ and\
  \bibinfo {author} {\bibfnamefont {O.}~\bibnamefont {Shishkina}},\ }\bibfield
  {title} {\bibinfo {title} {{Connecting wall modes and boundary zonal flows in
  rotating Rayleigh-B{\'{e}}nard convection}},\ }\href
  {https://doi.org/10.1103/PHYSREVFLUIDS.7.L011501/FIGURES/5/MEDIUM} {\bibfield
   {journal} {\bibinfo  {journal} {Phys. Rev. Fluids}\ }\textbf {\bibinfo
  {volume} {7}},\ \bibinfo {pages} {L011501} (\bibinfo {year}
  {2022})}\BibitemShut {NoStop}%
\bibitem [{\citenamefont {Madonia}\ \emph {et~al.}(2023)\citenamefont
  {Madonia}, \citenamefont {{Aguirre Guzm{\'{a}}n}}, \citenamefont {Clercx},\
  and\ \citenamefont {Kunnen}}]{Madonia2023}%
  \BibitemOpen
  \bibfield  {author} {\bibinfo {author} {\bibfnamefont {M.}~\bibnamefont
  {Madonia}}, \bibinfo {author} {\bibfnamefont {A.~J.}\ \bibnamefont {{Aguirre
  Guzm{\'{a}}n}}}, \bibinfo {author} {\bibfnamefont {H.~J.~H.}\ \bibnamefont
  {Clercx}},\ and\ \bibinfo {author} {\bibfnamefont {R.~P.~J.}\ \bibnamefont
  {Kunnen}},\ }\bibfield  {title} {\bibinfo {title} {{Reynolds number scaling
  and energy spectra in geostrophic convection}},\ }\href
  {https://doi.org/10.1017/jfm.2023.326} {\bibfield  {journal} {\bibinfo
  {journal} {J. Fluid Mech.}\ }\textbf {\bibinfo {volume} {962}},\ \bibinfo
  {pages} {A36} (\bibinfo {year} {2023})}\BibitemShut {NoStop}%
\bibitem [{\citenamefont {Favier}\ \emph {et~al.}(2014)\citenamefont {Favier},
  \citenamefont {Silvers},\ and\ \citenamefont {Proctor}}]{Favier2014}%
  \BibitemOpen
  \bibfield  {author} {\bibinfo {author} {\bibfnamefont {B.}~\bibnamefont
  {Favier}}, \bibinfo {author} {\bibfnamefont {L.~J.}\ \bibnamefont
  {Silvers}},\ and\ \bibinfo {author} {\bibfnamefont {M.~R.~E.}\ \bibnamefont
  {Proctor}},\ }\bibfield  {title} {\bibinfo {title} {{Inverse cascade and
  symmetry breaking in rapidly rotating Boussinesq convection}},\ }\href
  {https://doi.org/10.1063/1.4895131} {\bibfield  {journal} {\bibinfo
  {journal} {Phys. Fluids}\ }\textbf {\bibinfo {volume} {26}},\ \bibinfo
  {pages} {096605} (\bibinfo {year} {2014})}\BibitemShut {NoStop}%
\bibitem [{\citenamefont {Guervilly}\ \emph {et~al.}(2014)\citenamefont
  {Guervilly}, \citenamefont {Hughes},\ and\ \citenamefont
  {Jones}}]{Guervilly2014}%
  \BibitemOpen
  \bibfield  {author} {\bibinfo {author} {\bibfnamefont {C.}~\bibnamefont
  {Guervilly}}, \bibinfo {author} {\bibfnamefont {D.~W.}\ \bibnamefont
  {Hughes}},\ and\ \bibinfo {author} {\bibfnamefont {C.~A.}\ \bibnamefont
  {Jones}},\ }\bibfield  {title} {\bibinfo {title} {{Large-scale vortices in
  rapidly rotating Rayleigh-B{\'{e}}nard convection}},\ }\href
  {https://doi.org/10.1017/jfm.2014.542} {\bibfield  {journal} {\bibinfo
  {journal} {J. Fluid Mech.}\ }\textbf {\bibinfo {volume} {758}},\ \bibinfo
  {pages} {407\textendash435} (\bibinfo {year} {2014})}\BibitemShut {NoStop}%
\bibitem [{\citenamefont {{Aguirre Guzm{\'{a}}n}}\ \emph
  {et~al.}(2020)\citenamefont {{Aguirre Guzm{\'{a}}n}}, \citenamefont
  {Madonia}, \citenamefont {Cheng}, \citenamefont {Ostilla-M{\'{o}}nico},
  \citenamefont {Clercx},\ and\ \citenamefont {Kunnen}}]{Guzman2020}%
  \BibitemOpen
  \bibfield  {author} {\bibinfo {author} {\bibfnamefont {A.~J.}\ \bibnamefont
  {{Aguirre Guzm{\'{a}}n}}}, \bibinfo {author} {\bibfnamefont {M.}~\bibnamefont
  {Madonia}}, \bibinfo {author} {\bibfnamefont {J.~S.}\ \bibnamefont {Cheng}},
  \bibinfo {author} {\bibfnamefont {R.}~\bibnamefont {Ostilla-M{\'{o}}nico}},
  \bibinfo {author} {\bibfnamefont {H.~J.~H.}\ \bibnamefont {Clercx}},\ and\
  \bibinfo {author} {\bibfnamefont {R.~P.~J.}\ \bibnamefont {Kunnen}},\
  }\bibfield  {title} {\bibinfo {title} {{Competition between Ekman Plumes and
  Vortex Condensates in Rapidly Rotating Thermal Convection}},\ }\href
  {https://doi.org/10.1103/PhysRevLett.125.214501} {\bibfield  {journal}
  {\bibinfo  {journal} {Phys. Rev. Lett.}\ }\textbf {\bibinfo {volume} {125}},\
  \bibinfo {pages} {214501} (\bibinfo {year} {2020})}\BibitemShut {NoStop}%
\bibitem [{\citenamefont {Julien}\ \emph {et~al.}(2012)\citenamefont {Julien},
  \citenamefont {Rubio}, \citenamefont {Grooms},\ and\ \citenamefont
  {Knobloch}}]{Julien2012}%
  \BibitemOpen
  \bibfield  {author} {\bibinfo {author} {\bibfnamefont {K.}~\bibnamefont
  {Julien}}, \bibinfo {author} {\bibfnamefont {A.~M.}\ \bibnamefont {Rubio}},
  \bibinfo {author} {\bibfnamefont {I.}~\bibnamefont {Grooms}},\ and\ \bibinfo
  {author} {\bibfnamefont {E.}~\bibnamefont {Knobloch}},\ }\bibfield  {title}
  {\bibinfo {title} {{Statistical and physical balances in low Rossby number
  Rayleigh–B{\'{e}}nard convection}},\ }\href
  {https://doi.org/10.1080/03091929.2012.696109} {\bibfield  {journal}
  {\bibinfo  {journal} {Geophys. Astrophys. Fluid Dyn.}\ }\textbf {\bibinfo
  {volume} {106}},\ \bibinfo {pages} {392\textendash428} (\bibinfo {year}
  {2012})}\BibitemShut {NoStop}%
\bibitem [{\citenamefont {Rubio}\ \emph {et~al.}(2014)\citenamefont {Rubio},
  \citenamefont {Julien}, \citenamefont {Knobloch},\ and\ \citenamefont
  {Weiss}}]{Rubio2014}%
  \BibitemOpen
  \bibfield  {author} {\bibinfo {author} {\bibfnamefont {A.~M.}\ \bibnamefont
  {Rubio}}, \bibinfo {author} {\bibfnamefont {K.}~\bibnamefont {Julien}},
  \bibinfo {author} {\bibfnamefont {E.}~\bibnamefont {Knobloch}},\ and\
  \bibinfo {author} {\bibfnamefont {J.~B.}\ \bibnamefont {Weiss}},\ }\bibfield
  {title} {\bibinfo {title} {{Upscale Energy Transfer in Three-Dimensional
  Rapidly Rotating Turbulent Convection}},\ }\href
  {https://doi.org/10.1103/PhysRevLett.112.144501} {\bibfield  {journal}
  {\bibinfo  {journal} {Phys. Rev. Lett.}\ }\textbf {\bibinfo {volume} {112}},\
  \bibinfo {pages} {144501} (\bibinfo {year} {2014})}\BibitemShut {NoStop}%
\bibitem [{\citenamefont {de~Wit}\ \emph {et~al.}(2022)\citenamefont {de~Wit},
  \citenamefont {{Aguirre Guzm{\'a}n}}, \citenamefont {Clercx},\ and\
  \citenamefont {Kunnen}}]{DeWit2022}%
  \BibitemOpen
  \bibfield  {author} {\bibinfo {author} {\bibfnamefont {X.~M.}\ \bibnamefont
  {de~Wit}}, \bibinfo {author} {\bibfnamefont {A.~J.}\ \bibnamefont {{Aguirre
  Guzm{\'a}n}}}, \bibinfo {author} {\bibfnamefont {H.~J.~H.}\ \bibnamefont
  {Clercx}},\ and\ \bibinfo {author} {\bibfnamefont {R.~P.~J.}\ \bibnamefont
  {Kunnen}},\ }\bibfield  {title} {\bibinfo {title} {{Discontinuous transitions
  towards vortex condensates in buoyancy-driven rotating turbulence}},\ }\href
  {https://doi.org/10.1017/jfm.2022.90} {\bibfield  {journal} {\bibinfo
  {journal} {J. Fluid Mech.}\ }\textbf {\bibinfo {volume} {936}},\ \bibinfo
  {pages} {A43} (\bibinfo {year} {2022})}\BibitemShut {NoStop}%
\bibitem [{\citenamefont {Chandrasekhar}(1961)}]{Chandrasekhar1961}%
  \BibitemOpen
  \bibfield  {author} {\bibinfo {author} {\bibfnamefont {S.}~\bibnamefont
  {Chandrasekhar}},\ }\href@noop {} {\emph {\bibinfo {title} {{Hydrodynamic and
  Hydromagnetic Stability}}}}\ (\bibinfo  {publisher} {Oxford University
  Press},\ \bibinfo {year} {1961})\BibitemShut {NoStop}%
\bibitem [{\citenamefont {Verzicco}\ and\ \citenamefont
  {Orlandi}(1996)}]{Verzicco1996}%
  \BibitemOpen
  \bibfield  {author} {\bibinfo {author} {\bibfnamefont {R.}~\bibnamefont
  {Verzicco}}\ and\ \bibinfo {author} {\bibfnamefont {P.}~\bibnamefont
  {Orlandi}},\ }\bibfield  {title} {\bibinfo {title} {{A finite-difference
  scheme for three-dimensional incompressible flows in cylindrical
  coordinates}},\ }\href {https://doi.org/10.1006/jcph.1996.0033} {\bibfield
  {journal} {\bibinfo  {journal} {J. of Comput. Phys.}\ }\textbf {\bibinfo
  {volume} {123}},\ \bibinfo {pages} {402\textendash414} (\bibinfo {year}
  {1996})}\BibitemShut {NoStop}%
\bibitem [{\citenamefont {Verzicco}\ and\ \citenamefont
  {Camussi}(2003)}]{Verzicco2003}%
  \BibitemOpen
  \bibfield  {author} {\bibinfo {author} {\bibfnamefont {R.}~\bibnamefont
  {Verzicco}}\ and\ \bibinfo {author} {\bibfnamefont {R.}~\bibnamefont
  {Camussi}},\ }\bibfield  {title} {\bibinfo {title} {{Numerical experiments on
  strongly turbulent thermal convection in a slender cylindrical cell}},\
  }\href {https://doi.org/10.1017/S0022112002003063} {\bibfield  {journal}
  {\bibinfo  {journal} {J. Fluid Mech.}\ }\textbf {\bibinfo {volume} {477}},\
  \bibinfo {pages} {19\textendash49} (\bibinfo {year} {2003})}\BibitemShut
  {NoStop}%
\bibitem [{\citenamefont {Herrmann}\ and\ \citenamefont
  {Busse}(1993)}]{Herrmann1993}%
  \BibitemOpen
  \bibfield  {author} {\bibinfo {author} {\bibfnamefont {J.}~\bibnamefont
  {Herrmann}}\ and\ \bibinfo {author} {\bibfnamefont {F.~H.}\ \bibnamefont
  {Busse}},\ }\bibfield  {title} {\bibinfo {title} {{Asymptotic theory of
  wall-attached convection in a rotating fluid layer}},\ }\href
  {https://doi.org/10.1017/S0022112093002447} {\bibfield  {journal} {\bibinfo
  {journal} {J. Fluid Mech.}\ }\textbf {\bibinfo {volume} {255}},\ \bibinfo
  {pages} {183\textendash194} (\bibinfo {year} {1993})}\BibitemShut {NoStop}%
\bibitem [{\citenamefont {Zhang}\ and\ \citenamefont {Liao}(2009)}]{Zhang2009}%
  \BibitemOpen
  \bibfield  {author} {\bibinfo {author} {\bibfnamefont {K.}~\bibnamefont
  {Zhang}}\ and\ \bibinfo {author} {\bibfnamefont {X.}~\bibnamefont {Liao}},\
  }\bibfield  {title} {\bibinfo {title} {{The onset of convection in rotating
  circular cylinders with experimental boundary conditions}},\ }\href
  {https://doi.org/10.1017/S002211200800517X} {\bibfield  {journal} {\bibinfo
  {journal} {J. Fluid Mech.}\ }\textbf {\bibinfo {volume} {622}},\ \bibinfo
  {pages} {63\textendash73} (\bibinfo {year} {2009})}\BibitemShut {NoStop}%
\bibitem [{\citenamefont {Zhang}\ and\ \citenamefont {Liao}(2017)}]{Zhang2017}%
  \BibitemOpen
  \bibfield  {author} {\bibinfo {author} {\bibfnamefont {K.}~\bibnamefont
  {Zhang}}\ and\ \bibinfo {author} {\bibfnamefont {X.}~\bibnamefont {Liao}},\
  }\href {https://doi.org/10.1017/9781139024853} {\emph {\bibinfo {title}
  {{Theory and modeling of rotating fluids}}}}\ (\bibinfo  {publisher}
  {Cambridge University Press},\ \bibinfo {year} {2017})\BibitemShut {NoStop}%
\bibitem [{\citenamefont {Ecke}\ \emph {et~al.}(1992)\citenamefont {Ecke},
  \citenamefont {Zhong},\ and\ \citenamefont {Knobloch}}]{Ecke1992}%
  \BibitemOpen
  \bibfield  {author} {\bibinfo {author} {\bibfnamefont {R.~E.}\ \bibnamefont
  {Ecke}}, \bibinfo {author} {\bibfnamefont {F.}~\bibnamefont {Zhong}},\ and\
  \bibinfo {author} {\bibfnamefont {E.}~\bibnamefont {Knobloch}},\ }\bibfield
  {title} {\bibinfo {title} {{Hopf Bifurcation with Broken Reflection Symmetry
  in Rotating Rayleigh-B{\'{e}}nard Convection}},\ }\href
  {https://doi.org/10.1209/0295-5075/19/3/005} {\bibfield  {journal} {\bibinfo
  {journal} {Europhys. Lett.}\ }\textbf {\bibinfo {volume} {19}},\ \bibinfo
  {pages} {177} (\bibinfo {year} {1992})}\BibitemShut {NoStop}%
\bibitem [{\citenamefont {Goldstein}\ \emph {et~al.}(1993)\citenamefont
  {Goldstein}, \citenamefont {Knobloch}, \citenamefont {Mercader},\ and\
  \citenamefont {Net}}]{Goldstein1993}%
  \BibitemOpen
  \bibfield  {author} {\bibinfo {author} {\bibfnamefont {H.}~\bibnamefont
  {Goldstein}}, \bibinfo {author} {\bibfnamefont {E.}~\bibnamefont {Knobloch}},
  \bibinfo {author} {\bibfnamefont {I.}~\bibnamefont {Mercader}},\ and\
  \bibinfo {author} {\bibfnamefont {M.}~\bibnamefont {Net}},\ }\bibfield
  {title} {\bibinfo {title} {Convection in a rotating cylinder. {P}art 1.
  {L}inear theory for moderate {P}randtl numbers},\ }\href
  {https://doi.org/10.1017/S0022112093000928} {\bibfield  {journal} {\bibinfo
  {journal} {J. Fluid Mech.}\ }\textbf {\bibinfo {volume} {248}},\ \bibinfo
  {pages} {583\textendash604} (\bibinfo {year} {1993})}\BibitemShut {NoStop}%
\bibitem [{\citenamefont {Zhong}\ \emph {et~al.}(1993)\citenamefont {Zhong},
  \citenamefont {Ecke},\ and\ \citenamefont {Steinberg}}]{Zhong1993}%
  \BibitemOpen
  \bibfield  {author} {\bibinfo {author} {\bibfnamefont {F.}~\bibnamefont
  {Zhong}}, \bibinfo {author} {\bibfnamefont {R.~E.}\ \bibnamefont {Ecke}},\
  and\ \bibinfo {author} {\bibfnamefont {V.}~\bibnamefont {Steinberg}},\
  }\bibfield  {title} {\bibinfo {title} {{Rotating Rayleigh–B{\'{e}}nard
  convection: asymmetric modes and vortex states}},\ }\href
  {https://doi.org/10.1017/S0022112093001119} {\bibfield  {journal} {\bibinfo
  {journal} {J. Fluid Mech.}\ }\textbf {\bibinfo {volume} {249}},\ \bibinfo
  {pages} {135\textendash159} (\bibinfo {year} {1993})}\BibitemShut {NoStop}%
\bibitem [{\citenamefont {Kunnen}\ \emph {et~al.}(2011)\citenamefont {Kunnen},
  \citenamefont {Stevens}, \citenamefont {Overkamp}, \citenamefont {Sun},
  \citenamefont {van Heijst},\ and\ \citenamefont {Clercx}}]{Kunnen2011}%
  \BibitemOpen
  \bibfield  {author} {\bibinfo {author} {\bibfnamefont {R.~P.~J.}\
  \bibnamefont {Kunnen}}, \bibinfo {author} {\bibfnamefont {R.~J. A.~M.}\
  \bibnamefont {Stevens}}, \bibinfo {author} {\bibfnamefont {J.}~\bibnamefont
  {Overkamp}}, \bibinfo {author} {\bibfnamefont {C.}~\bibnamefont {Sun}},
  \bibinfo {author} {\bibfnamefont {G.~J.~F.}\ \bibnamefont {van Heijst}},\
  and\ \bibinfo {author} {\bibfnamefont {H.~J.~H.}\ \bibnamefont {Clercx}},\
  }\bibfield  {title} {\bibinfo {title} {{The role of Stewartson and Ekman
  layers in turbulent rotating Rayleigh–B{\'{e}}nard convection}},\ }\href
  {https://doi.org/10.1017/JFM.2011.383} {\bibfield  {journal} {\bibinfo
  {journal} {J. Fluid Mech.}\ }\textbf {\bibinfo {volume} {688}},\ \bibinfo
  {pages} {422\textendash442} (\bibinfo {year} {2011})}\BibitemShut {NoStop}%
\bibitem [{\citenamefont {Kunnen}\ \emph {et~al.}(2013)\citenamefont {Kunnen},
  \citenamefont {Clercx},\ and\ \citenamefont {van Heijst}}]{Kunnen2013}%
  \BibitemOpen
  \bibfield  {author} {\bibinfo {author} {\bibfnamefont {R.~P.~J.}\
  \bibnamefont {Kunnen}}, \bibinfo {author} {\bibfnamefont {H.~J.~H.}\
  \bibnamefont {Clercx}},\ and\ \bibinfo {author} {\bibfnamefont {G.~J.~F.}\
  \bibnamefont {van Heijst}},\ }\bibfield  {title} {\bibinfo {title} {{The
  structure of sidewall boundary layers in confined rotating
  Rayleigh-B{\'{e}}nard convection}},\ }\href
  {https://doi.org/10.1017/jfm.2013.285} {\bibfield  {journal} {\bibinfo
  {journal} {J. Fluid Mech.}\ }\textbf {\bibinfo {volume} {727}},\ \bibinfo
  {pages} {509\textendash532} (\bibinfo {year} {2013})}\BibitemShut {NoStop}%
\bibitem [{\citenamefont {Liao}\ \emph {et~al.}(2006)\citenamefont {Liao},
  \citenamefont {Zhang},\ and\ \citenamefont {Chang}}]{Liao2006}%
  \BibitemOpen
  \bibfield  {author} {\bibinfo {author} {\bibfnamefont {X.}~\bibnamefont
  {Liao}}, \bibinfo {author} {\bibfnamefont {K.}~\bibnamefont {Zhang}},\ and\
  \bibinfo {author} {\bibfnamefont {Y.}~\bibnamefont {Chang}},\ }\bibfield
  {title} {\bibinfo {title} {{Convection in rotating annular channels heated
  from below. Part 1. Linear stability and weakly nonlinear mean flows}},\
  }\href {https://doi.org/10.1080/03091920500291805} {\bibfield  {journal}
  {\bibinfo  {journal} {Geophys. Astrophys. Fluid Dyn.}\ }\textbf {\bibinfo
  {volume} {99}},\ \bibinfo {pages} {445\textendash465} (\bibinfo {year}
  {2006})}\BibitemShut {NoStop}%
\bibitem [{\citenamefont {Stewartson}(1957)}]{Stewartson1957}%
  \BibitemOpen
  \bibfield  {author} {\bibinfo {author} {\bibfnamefont {K.}~\bibnamefont
  {Stewartson}},\ }\bibfield  {title} {\bibinfo {title} {{On almost rigid
  rotations}},\ }\href {https://doi.org/10.1017/S0022112057000452} {\bibfield
  {journal} {\bibinfo  {journal} {J. Fluid Mech.}\ }\textbf {\bibinfo {volume}
  {3}},\ \bibinfo {pages} {17\textendash26} (\bibinfo {year}
  {1957})}\BibitemShut {NoStop}%
\bibitem [{\citenamefont {{Aguirre Guzm{\'{a}}n}}\ \emph
  {et~al.}(2021)\citenamefont {{Aguirre Guzm{\'{a}}n}}, \citenamefont
  {Madonia}, \citenamefont {Cheng}, \citenamefont {Ostilla-M{\'{o}}nico},
  \citenamefont {Clercx},\ and\ \citenamefont {Kunnen}}]{AguirreGuzman2021}%
  \BibitemOpen
  \bibfield  {author} {\bibinfo {author} {\bibfnamefont {A.~J.}\ \bibnamefont
  {{Aguirre Guzm{\'{a}}n}}}, \bibinfo {author} {\bibfnamefont {M.}~\bibnamefont
  {Madonia}}, \bibinfo {author} {\bibfnamefont {J.~S.}\ \bibnamefont {Cheng}},
  \bibinfo {author} {\bibfnamefont {R.}~\bibnamefont {Ostilla-M{\'{o}}nico}},
  \bibinfo {author} {\bibfnamefont {H.~J.~H.}\ \bibnamefont {Clercx}},\ and\
  \bibinfo {author} {\bibfnamefont {R.~P.~J.}\ \bibnamefont {Kunnen}},\
  }\bibfield  {title} {\bibinfo {title} {{Force balance in rapidly rotating
  Rayleigh–B{\'{e}}nard convection}},\ }\href
  {https://doi.org/10.1017/JFM.2021.802} {\bibfield  {journal} {\bibinfo
  {journal} {J. Fluid Mech.}\ }\textbf {\bibinfo {volume} {928}},\ \bibinfo
  {pages} {A16} (\bibinfo {year} {2021})}\BibitemShut {NoStop}%
\end{thebibliography}%

\end{document}